\documentclass[usenatbib,useAMS]{mnras}
\usepackage{graphicx}	
\usepackage{amsmath}	
\usepackage{amssymb}	
\usepackage{multicol}
\usepackage{bm}		
\usepackage[english]{babel}
\usepackage{epstopdf}

\newcommand{\mpchi}{\,h^{-1}{\rm {Mpc}}}
\newcommand{\kms}{\,{\rm {km\, s^{-1}}}}
\newcommand{\msun}{\,h^{-1}{\rm M_{\sun}}}

\title[Redshift-space Clustering]{Redshift-Space Clustering of SDSS Galaxies --- Luminosity Dependence, Halo Occupation
Distribution, and Velocity Bias}

\author[Guo et al.]{\parbox{\textwidth}{
Hong Guo$^{1,2}$\thanks{E-mail: guohong@shao.ac.cn}, Zheng Zheng$^{2}$, Idit
Zehavi$^{3,4}$, Peter S. Behroozi$^{5}$, Chia-Hsun Chuang$^{6}$, Johan
Comparat$^{6,7}$, Ginevra Favole$^{6,8}$, Stefan Gottloeber$^{9}$, Anatoly
Klypin$^{10}$, Francisco Prada$^{6,8,11}$, David H. Weinberg$^{12,13}$, and
Gustavo Yepes$^{7}$}
\vspace*{6pt} \\
$^{1}$ Shanghai Astronomical Observatory, Chinese Academy of Sciences, Shanghai 200030, China\\
$^{2}$ Department of Physics and Astronomy, University of Utah, UT 84112, USA\\
$^{3}$ Department of Astronomy, Case Western Reserve University, OH 44106, USA\\
$^{4}$ Institute for Computational Cosmology, Department of Physics,
University of Durham, South Road, Durham, DH1 3LE,
UK\\
$^{5}$ Space Telescope Science Institute, Baltimore, MD 21218, USA\\
$^{6}$ Instituto de F\'{\i}sica Te\'orica, (UAM/CSIC), Universidad Aut\'onoma
de Madrid,  Cantoblanco, E-28049 Madrid,
Spain \\
$^{7}$ Departamento de F{\'i}sica Te{\'o}rica,  Universidad Aut{\'o}noma de Madrid, Cantoblanco, 28049, Madrid, Spain\\
$^{8}$ Campus of International Excellence UAM+CSIC, Cantoblanco, E-28049 Madrid, Spain\\
$^{9}$ Leibniz-Institut fur Astrophysik (AIP), An der Sternwarte 16, D-14482 Potsdam, Germany\\
$^{10}$ Astronomy Department, New Mexico State University, MSC 4500, PO Box 30001, Las Cruces, NM, 880003-8001, USA\\
$^{11}$ Instituto de Astrof\'{\i}sica de Andaluc\'{\i}a (CSIC), Glorieta de
la Astronom\'{\i}a, E-18080 Granada, Spain
\\
$^{12}$ Department of Astronomy, Ohio State University, Columbus, OH 43210, USA\\
$^{13}$ Center for Cosmology and Astro-Particle Physics, Ohio State
University, Columbus, OH 43210, USA}

\begin{document}
\label{firstpage} \pagerange{\pageref{firstpage}--\pageref{lastpage}}
\maketitle

\begin{abstract}
We present the measurements and modelling of the small-to-intermediate scale
($\sim 0.1$--$25\mpchi$) projected and three-dimensional (3D) redshift-space
two-point correlation functions (2PCFs) of local galaxies in the Sloan
Digital Sky Survey (SDSS) Data Release 7. We find a clear dependence of
galaxy clustering on luminosity in both projected and redshift spaces,
generally being stronger for more luminous samples. The measurements are
successfully interpreted within the halo occupation distribution (HOD)
framework with central and satellite velocity bias parameters to describe
galaxy kinematics inside haloes and to model redshift-space distortion (RSD)
effects. In agreement with previous studies, we find that more luminous
galaxies reside in more massive haloes. Including the redshift-space 2PCFs
helps tighten the HOD constraints. Moreover, we find that luminous central
galaxies are not at rest at the halo centres, with the velocity dispersion
about 30\% that of the dark matter. Such a relative motion may reflect the
consequence of galaxy and halo mergers, and we find that central galaxies in
lower mass haloes tend to be more relaxed with respect to their host haloes.
The motion of satellite galaxies in luminous samples is consistent with their
following that of the dark matter. For faint samples, satellites tends to
have slower motion, with velocity dispersion inside haloes about 85\% that of
the dark matter. We discuss possible applications of the velocity bias
constraints on studying galaxy evolution and cosmology. In the appendix, we
characterize the distribution of galaxy redshift measurement errors, which is
well described by a Gaussian-convolved double exponential distribution.
\end{abstract}


\section{Introduction}
The three-dimensional galaxy distribution in our Universe can be probed
through the large-scale galaxy redshift surveys, such as the Sloan Digital
Sky Survey \citep[SDSS;][]{York00}. The angular positions of the galaxies can
be accurately measured in photometric observations, while the radial
positions are usually obtained from the observed galaxy redshifts. However,
the radial distances derived from the redshifts differ from the real
positions of galaxies due to the existence of the galaxy peculiar velocities,
which is usually referred to as the redshift-space distortion (RSD) effect.
Although the RSD prevents us from measuring the true galaxy distribution, it
also provides valuable information about the galaxy kinematics in dark matter
haloes.

A commonly used statistic for analysing the galaxy distribution is the
three-dimensional (3D) two-point correlation function (2PCF),
$\xi(r_p,r_{\rm\pi})$, where $r_p$ and $r_{\rm\pi}$ are the transverse and
line-of-sight (LOS) separations of galaxy pairs, respectively \citep[see
e.g.][]{Zehavi05,Zehavi11,Li06,Guo13}. To minimise the effect of RSD, the
traditional way of probing the real-space galaxy distribution is through the
projected 2PCF, $w_p(r_p)$, by integrating $\xi(r_p,r_{\rm\pi})$ along the
LOS direction. Since $w_p(r_p)$ is insensitive to the galaxy peculiar
velocities, it only probes the galaxy spatial distribution. To fully
understand the galaxy phase-space distribution, we need to model the
redshift-space clustering of galaxies.

In contemporary galaxy formation and evolution models, galaxies form and
evolve in dark matter haloes. The galaxy distribution in the universe can
then be studied with the halo model, through the distribution of galaxies
within the haloes and the halo distributions in the large-scale structure of
the universe \citep[see e.g.][and references therein]{Cooray02}. Since the
distribution of dark matter haloes is well understood using analytic models
and numerical $N$-body simulations \cite[e.g.,][]{Mo96,Springel05,Klypin14},
the key component of the models is the connection between galaxies and dark
matter haloes, such as the framework of halo occupation distribution (HOD) or
the closely related conditional luminosity function
\citep[e.g.][]{Jing98,Peacock00,Seljak00,Scoccimarro01,Berlind02,Yang03,Zheng05,Guo14,Skibba15}.
The HOD describes the probability distribution $P(N|M)$ of having $N$
galaxies of a given type in a dark matter halo of virial mass $M$. The
probability distribution $P(N|M)$, together with the spatial and velocity
distributions of galaxies inside haloes, is crucial to interpret and
understand the real- and redshift-space clustering of galaxies. The
observationally inferred HOD can help us test and constrain galaxy formation
models.

A large fraction of the information contents on the HOD (like the occupation
function and galaxy kinematics and spatial distribution inside haloes) are
contained in the small-scale clustering of galaxies. However, both measuring
and modelling the galaxy distribution are nontrivial in the small-scale
nonlinear regime. On the observational side, in fibre-fed spectrograph
surveys as in the SDSS, the hardware limit that two fibres on the same plate
cannot be placed closer than an angular separation of $55''$
\citep{Blanton03} significantly hinders the number of close galaxy pairs on
small scales. Fortunately, this fibre collision effect can be accurately
corrected using the method of \cite*{Guo12}, by taking advantage of the
recovered redshifts of collided galaxies in the plate-overlap regions. On the
theory side, modelling the galaxy distributions on small scales is also
difficult, especially in redshift space due to the lack of understanding of
the galaxy phase-space distribution. This problem can be alleviated with the
help of high-resolution $N$-body dark matter simulations.

Recently, \cite{Guo15a} (hereafter G15) measured and modelled the luminous
red galaxy (LRG) distribution in the SDSS-III Baryon Oscillation
Spectroscopic Survey \citep[BOSS;][]{Eisenstein11,Dawson13} at redshift
$z{\sim}0.5$, and found that central galaxies are not at rest at the halo
centres and the satellite galaxies move more slowly than the dark matter. The
difference in galaxy and matter velocity distributions is dubbed as velocity
bias. In this paper, we follow the method of G15 and infer the phase-space
distribution of galaxies in the local universe through modelling the
redshift-space galaxy 2PCFs. We improve the model of G15 and also incorporate
a more accurate redshift error model shown in Appendix~\ref{app:zerr}. In
Section 2, we describe the data, the galaxy samples, and the redshift-space
2PCF measurements. We introduce our modelling method in Section 3. The
constraints on the occupation function and galaxy phase-space distribution
are presented in Section 4. Finally, we summarize our results in Section 5.

Throughout the paper, for the measurements we assume a spatially flat
$\Lambda$ cold dark matter ($\Lambda$CDM) cosmology, with $\Omega_m=0.307$
and $h=0.678$, which is adopted in the MultiDark simulation (and model) we
use, and consistent with the constraints from Planck \citep{Planck14}.

\section{Data and Measurements}\label{sec:data}
\begin{figure}
\includegraphics[width=0.45\textwidth]{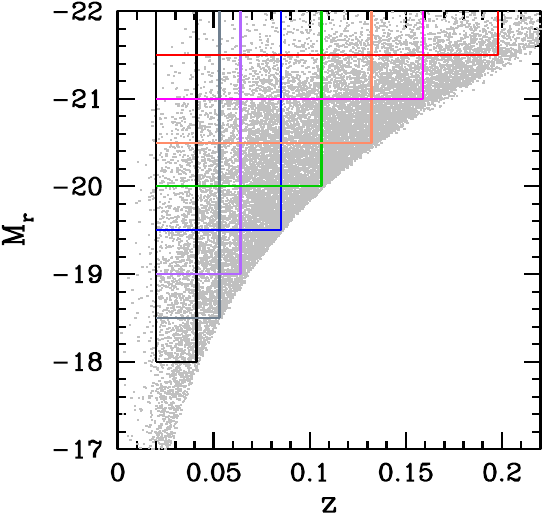}
\caption{Construction of volume-limited galaxy samples with different luminosity thresholds. The dots show the
distribution of SDSS galaxies as a function of redshift $z$ and $r$-band absolute magnitude $M_r$. The different colour
lines delineate the selection cuts for different luminosity-threshold samples defined in Table~\ref{tab:sample}.}
\label{fig:mlz}
\end{figure}
For the purpose of studying the galaxy distribution in the local universe, we
use the galaxy sample of the New York University Value-Added Galaxy Catalog
\citep[NYU-VAGC;][]{Blanton05b}, which is constructed from the SDSS Data
Release 7 Main galaxy sample \citep{Abazajian09}. The sample covers an
effective area of about $7300\deg^2$, with galaxies selected using an
$r$-band Petrosian magnitude limit of $r<17.77$. The magnitudes in the
catalog are $K$--corrected and passively evolving to the median redshift of
$z=0.1$ \citep{Blanton03b}. To properly measure and model the galaxy
clustering and its dependence on galaxy luminosity, we construct
volume-limited samples with different luminosity thresholds. We impose a
minimum redshift of $z=0.02$. The selection cuts for the different samples
are shown in Fig.~\ref{fig:mlz}. Table~\ref{tab:sample} provides the
corresponding sample information, including the average number density and
the volume of each sample. Overall, the samples are similar to those
constructed in \cite{Zehavi11}.

We measure the 3D redshift-space 2PCFs $\xi(r_p,r_{\rm\pi})$ of the SDSS DR7
Main galaxy sample through the Landy--Szalay estimator \citep{Landy93}. The
2PCF, $\xi(r_p,r_{\rm\pi})$, can be further integrated along the LOS to
reduce the effect of RSD. The resulting projected 2PCF $w_p(r_p)$
\citep{Davis83} is defined as
\begin{equation}
w_p(r_p)=2\int_0^\infty \xi(r_p,r_{\rm\pi})dr_{\rm\pi}. \label{eq:wp}
\end{equation}
In both the measurements and the model, we integrate $\xi(r_p,r_\pi)$ to
$r_{\pi,{\rm max}}=40\mpchi$ to obtain $w_p$. We adopt logarithmic $r_p$ bins
centred at $0.13$ to $20.48\mpchi$ with $\Delta\log r_p=0.2$, and linear
$r_{\rm\pi}$ bins from 0 to 40$\mpchi$ with $\Delta r_{\rm\pi}=2\mpchi$.
Denoting the relative redshift-space position of a pair of galaxies as
$\bmath{s}$, we also measure the redshift-space 2PCF in bins of $s$ and
$\mu$, with $s=\sqrt{r_p^2+r_{\rm\pi}^2}$ being the redshift-space separation
of galaxy pairs and $\mu$ being the cosine of the angle between $\bmath{s}$
and the LOS. The redshift-space 2PCF $\xi(s,\mu)$ can be expanded into
multipoles \citep{Hamilton92},
\begin{equation}
\xi(s,\mu)=\sum_l\xi_l(s)P_l(\mu),
\end{equation}
where $P_l$ is the $l$-th order Legendre polynomial. The multipole moments
$\xi_l$ are usually used to characterize the redshift-space clustering (G15).
We focus on the measurements of the monopole ($\xi_0$), quadrupole ($\xi_2$),
and hexadecapole ($\xi_4$). For $s$, we adopt the same logarithmic bins as
$r_p$, while for $\mu$ we use linear bins from -1 to 1 with $\Delta\mu=0.05$.

In linear theory, the three multipole moments we adopt are
the only non-zero terms. While at any scales odd terms are zero by the
symmetry of $\xi(s,\mu)$, at the translinear or nonlinear scales explored in
this paper, higher-order even terms do exist. However, the information
content in the higher-order terms is minimal compared to the three main terms
($\xi_0$, $\xi_2$, and $\xi_4$), as they are highly correlated with these
lower-order terms. For example, \cite{Hikage14} explored the constraints on
the HOD by including multipoles of different orders, and found that including
the tetra-hexadecapole ($l=6$) in addition to multipoles with $l=$0, 2, and 4
leads to virtually no improvement in the constraints (see their Table~1 and
Fig.~3). We therefore limit our study to only $\xi_0$, $\xi_2$, and $\xi_4$,
in addition to $w_p$.

\begin{table}
\caption{Samples of different luminosity thresholds} \label{tab:sample}
\begin{tabular}{@{}lrrrr@{}}
\hline
$M_r^{\rm max}$  & $z_{\rm max}$ & $N$ & $n_g$ & $V$\\
\hline
$-18.0$  & 0.041 & 35649 & $31.42\pm4.03$ & $  1.19$\\
$-18.5$  & 0.053 & 57786 & $22.25\pm2.70$ & $  2.72$\\
$-19.0$  & 0.064 & 72484 & $15.66\pm2.06$ & $  4.87$\\
$-19.5$  & 0.085 & 125664 & $11.64\pm1.27$ & $ 11.44$\\
$-20.0$  & 0.106 & 131623 & $ 6.37\pm0.75$ & $ 22.00$\\
$-20.5$  & 0.132 & 122115 & $ 3.13\pm0.30$ & $ 41.82$\\
$-21.0$  & 0.159 & 76802 & $ 1.16\pm0.12$ & $ 71.74$\\
$-21.5$  & 0.198 & 35207 & $ 0.29\pm0.03$ & $134.65$\\
\hline
\end{tabular}
\medskip

The absolute magnitude is computed by assuming $h=1$. The minimum redshift of
all the luminosity threshold samples is $z_{\rm min}=0.02$. The total number
of galaxies $N$ in each sample is also displayed. The mean number density
$n_g$ is in units of $10^{-3}h^{3}{\rm {Mpc}}^{-3}$. The volume $V$ of each
sample is in units of $10^6h^{-3}{\rm {Mpc}}^3$.
\end{table}

We apply the method developed in \cite*{Guo12} to correct the fibre-collision
effect, enabling accurate measurements of the small-scale 2PCFs. For each
sample, the covariance matrix of the measurements is estimated from 400
jackknife samples \citep{Zehavi11,Guo15a}.

\begin{figure*}
\includegraphics[width=0.7\textwidth]{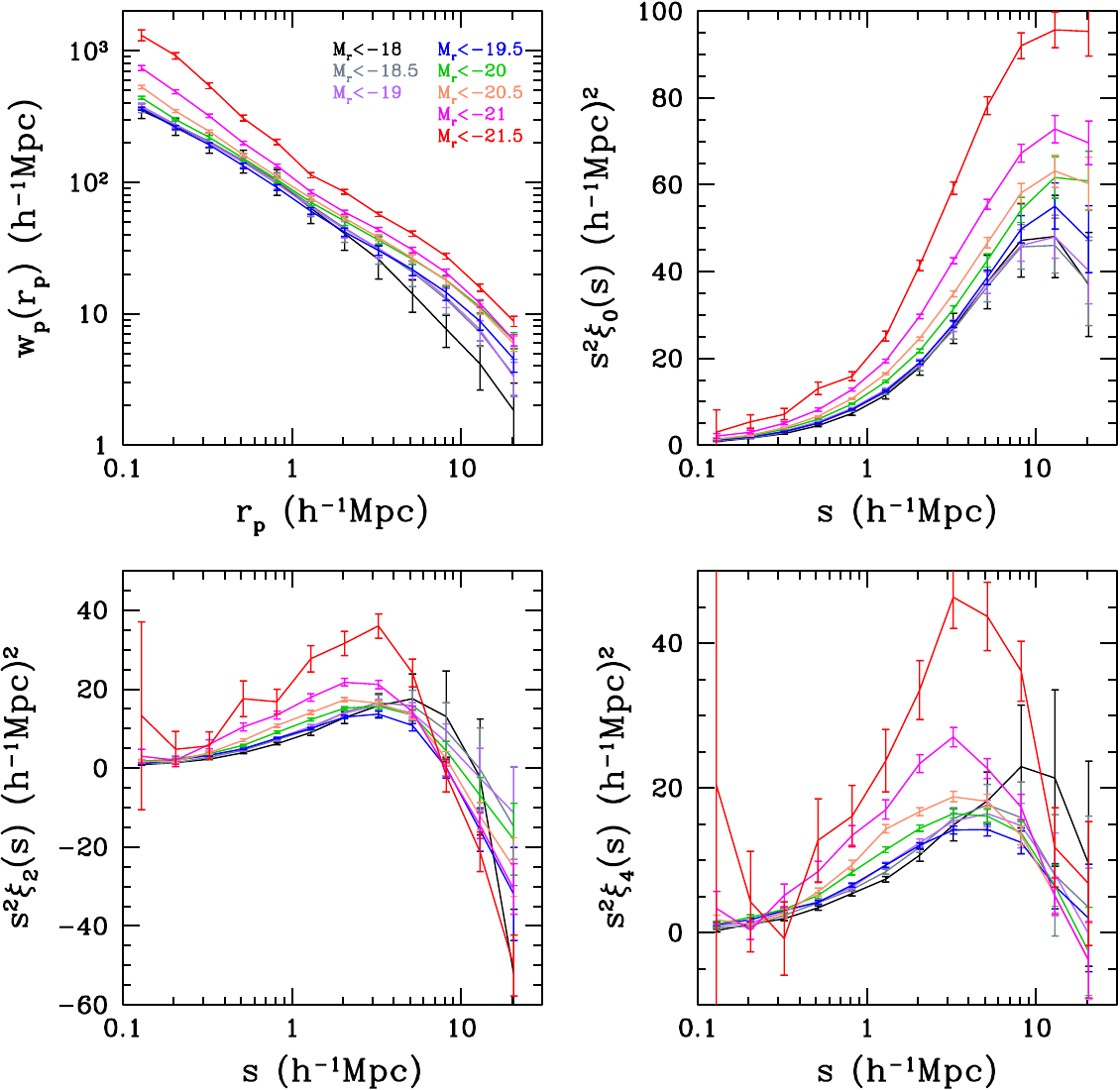}
\caption{Measured projected 2PCF $w_p(r_p)$ (top left), redshift-space
monopole $\xi_0(s)$ (top right), quadrupole $\xi_2(s)$ (bottom left), and
hexadecapole $\xi_4(s)$ (bottom right) for SDSS galaxies. The different
colour lines show the measurements for different luminosity threshold
samples. The errors are estimated from 400 jackknife samples.} \label{fig:wx}
\end{figure*}

We show in Fig.~\ref{fig:wx} the measured 2PCFs for the different luminosity
threshold samples. It is clear from the projected 2PCF $w_p(r_p)$ and the
redshift-space monopole $\xi_0(s)$ that more luminous galaxies have stronger
clustering amplitudes than their fainter counterparts, consistent with the
results of \cite{Zehavi11}. The measurements of the quadrupole $\xi_2(s)$ and
hexadecapole $\xi_4(s)$ are noisier, but the overall dependence on luminosity
is clear. For example, more luminous galaxies show a more positive quadrupole
and hexadecapole on small scales, indicating stronger Fingers-of-God (FOG)
effects. We jointly model all the 2PCF measurements in the following
sections.

\section{Simulation and Model}
\label{sec:simmod}
To accurately model the galaxy clustering under the HOD framework, we follow
the simulation-based model method developed in
\citet{Zheng15}, which is used in G15 to model redshift-space clustering of
BOSS galaxies. It tabulates properties of haloes in an $N$-body simulation
(e.g., halo mass functions, halo profiles, halo clustering) necessary for
computing galaxy 2PCFs. Given a set of HOD parameters, the model can then
accurately predict the galaxy 2PCFs. It is equivalent to assigning galaxies
to haloes in the simulation with the given set of HOD parameters and
measuring the 2PCFs from the resultant mock catalog. When
populating haloes with galaxies, the redshift-space distortion is applied in
a galaxy-by-galaxy manner by using the velocity information of galaxies and
haloes. Our modelling method does the same thing by using the galaxy velocity
distribution inside haloes and the redshift-space clustering of haloes (see
\citealt{Zheng15} for details). Our simulation-based modelling method works
more efficiently than directly populating the simulation with galaxies, since
it enables `populating galaxies'
 and `measuring the 2PCF in the mock' to be performed analytically.
The model is accurate since it automatically takes into account the effects
of halo exclusion, nonlinear growth, and scale-dependent halo bias by using
the halo catalogues in high-resolution simulations. In particular, it is well
suited to model the redshift-space galaxy clustering on small and
intermediate scales, for which an accurate analytic model is difficult to
develop \citep[e.g.][]{Tinker07,Reid11,WRW14,White15}.

The model we use in this paper is based on the MultiDark simulation of Planck
cosmology (MDPL; \citealt{Klypin14}), which is carried out with L-GADGET-2
code \citep{Springel05}. The cosmological parameters ($\Omega_m=0.307$,
$\Omega_b=0.048$, $h=0.678$, $n_s=0.96$, and $\sigma_8=0.823$) used in MDPL
are consistent with the recent results from Planck \citep{Planck14}. The
simulation has 3840$^{3}$ dark matter particles in a box of 1\,$h^{-1}$\,Gpc
(comoving) on a side, so the mass resolution is $1.51\times10^9\msun$, which
is about 6 times higher than the previous MultiDark run simulation (MDR1;
\citealt{Prada12}). The force resolution, i.e. the gravitational softening
length, is only $5h^{-1}$\,kpc (physical) at low redshifts, which enables us
to accurately model the clustering signals on very small scales. We use the
simulation output at $z=0$ to model all the luminosity threshold galaxy
samples in the NYU-VAGC. In principle, when modelling the
measurements, it is better to choose the simulation output to match the mean
redshift for each individual sample. This is certainly limited by the
available simulation outputs. On the other hand, given the small redshift
range of the SDSS Main galaxies, the effect of using one output (as we do) is
small. We have tested applying the $z=0.1$ simulation output for modelling
the data, and the inferred HOD parameters are consistent with those from the
default model built on the $z=0$ output.

Dark matter haloes are identified using the Rockstar phase-space halo finder
\citep{Behroozi13}, which is efficient and accurate in
finding the bound spherical structures from the density peaks in the phase
space \citep{Onions12,Knebe13}. Halo mass is defined from the given spherical
overdensities of a virial structure \citep{Bryan98}. Note that we do not
remove the unbound particles in the haloes, because the satellite galaxies in
the haloes can also be unbound. The halo positions, velocities and velocity
dispersions are calculated from all the particles in the haloes. In Rockstar,
the centre of each halo is computed from the average particle locations for
the inner friends-of-friends (FOF) subgroup that best minimizes the Poisson
error. Different from G15, who use the average velocity of the inner $25$ per
cent of halo particles as the halo core velocity, we define the halo velocity
as the average velocity of all particles in the halo, i.e. the centre-of-mass
velocity. The purpose of this definition is to make better comparisons with
the literature, and also make easier the application of our models to other
low-resolution simulations. The definition of the halo velocity is important
for comparing the results of galaxy velocity bias, since the halo core
velocity can have a substantial velocity offset from the halo bulk velocity
\citep{Behroozi13,Reid14}.

For the HOD modelling of the galaxy clustering, we follow the parametrization
of \cite*{Zheng07} by decomposing the contributions to the mean occupation
function $\langle N(M)\rangle$ of galaxies (i.e. the average number $N$ of
galaxies in a sample in haloes of mass $M$) into the central and satellite
components,
\begin{eqnarray}
\langle N(M)\rangle&=&\langle N_{\rm cen}(M)\rangle+\langle N_{\rm sat}(M)\rangle,\\
\langle N_{\rm cen}(M)\rangle&=&\frac{1}{2}\left[1+{\rm erf}\left(\frac{\log M-\log M_{\rm min}}{\sigma_{\log
M}}\right)\right], \label{eqn:Ncen}\\
\langle N_{\rm sat}(M)\rangle&=&\langle N_{\rm cen}(M)\rangle\left(\frac{M-M_0}{M_1^\prime}\right)^\alpha,
\label{eqn:Nsat}
\end{eqnarray}
where $M_{\rm min}$ describes the cutoff halo mass of the central galaxies
and $\sigma_{\log M}$ takes into account the scatter between the galaxy
luminosity and halo mass. The three parameters for the satellite galaxies are
the cutoff mass scale $M_0$, the normalization mass scale $M_1^\prime$ and
the power-law slope $\alpha$ at the high-mass end. In our model, we
implicitly assume that the halo hosting a satellite galaxy in a given
luminosity threshold sample also hosts a central galaxy from the same sample.
One derived parameter we have is $M_1$, the characteristic mass of haloes
hosting on average one satellite galaxy. In combination with the halo mass
function, the satellite fraction $f_{\rm sat}$ of the galaxies in the sample
can also be derived from the central and satellite mean occupation functions.

To model the redshift-space galaxy clustering, we need to
specify the phase-space (spatial and velocity) distribution of galaxies
inside haloes. We put the central galaxies at the halo centres and randomly
select dark matter particles inside haloes to represent the satellite
galaxies. When calculating the redshift-space clustering, we employ the
plane-parallel approximation and use the $\hat{z}$ direction in the
simulation as the LOS. The shift of a galaxy's position from real space to
redshift space in the $\hat{z}$ direction due to the RSD is then calculated
as $\Delta Z=v_{\hat{z}}(1+z)/H(z)$ with $z=0$, where $v_{\hat{z}}$ is the
LOS peculiar velocity of the galaxy.

As found by G15, the motion of galaxies can differ from that of dark matter.
We therefore introduce two velocity bias parameters ($\alpha_c$ and
$\alpha_s$) in the HOD model when describing the velocities of central and
satellite galaxies. Note that all the velocity quantities below refer to the
$\hat{z}$ (LOS) component for our modelling purpose, including the central
galaxy velocity $v_c$, satellite galaxy velocity $v_s$, halo velocity $v_h$,
and particle velocity dispersion $\sigma_v$ inside a given halo. We first
measure the LOS velocity dispersion $\sigma_v$ from all the dark matter
particles in the haloes. The central galaxy is not necessarily at rest with
respect to the host halo, and its velocity $v_c-v_h$ in the frame of the halo
is assumed to follow a Laplace distribution, in the form of
\begin{equation}
f(v_c-v_h)=\frac{1}{\sqrt{2}\sigma_c}\exp\left(-\frac{\sqrt{2}|v_c-v_h|}{\sigma_c}\right),
\end{equation}
where $v_h$ is the LOS centre-of-mass velocity of the halo,
$\sigma_c\equiv\alpha_c\sigma_v$ is the LOS central galaxy velocity
dispersion, and $\alpha_c$ is the central galaxy velocity bias,
characterizing the relative motion between the central galaxy and the host
halo. The use of Laplace distribution instead of the commonly-used Gaussian
distribution is motivated by the distribution of the velocities of brightest
cluster galaxies (BCGs) relative to satellites in Abell clusters
\citep{Lauer14}.

To allow for possible velocity offsets between the satellite galaxies and the
randomly-selected dark matter particles, we scale the velocity of the
satellite galaxies in the centre-of-mass frame of the halo by a satellite
velocity bias factor $\alpha_s$,
\begin{equation}
\label{eqn:alpha_s} v_s-v_h=\alpha_s(v_p-v_h),
\end{equation}
where $v_s$ and $v_p$ are the LOS velocities of the satellite galaxies and
the selected dark matter particles, respectively. Therefore, the LOS velocity
dispersion $\sigma_s$ of satellite galaxies in the haloes is
$\sigma_s=\alpha_s\sigma_v$ \citep[see, e.g.][]{Tinker07}. We note that even
though we only apply the LOS velocity bias in the above equations, the
velocity bias exists in all components of the galaxy velocities. However, for
the purpose of modelling the redshift-space clustering, only the LOS
component matters. In our fiducial HOD model, we assume a constant galaxy
velocity bias, good enough given the current data precision.

Except for the galaxy velocity bias, another ingredient that could affect the
galaxy LOS distribution in redshift space is the measurement error of the
SDSS galaxy redshifts. We show in Appendix~\ref{app:zerr} an accurate
modelling of the redshift errors from repeat observations of galaxy spectra
in the SDSS. We find that the additional velocity contribution introduced by
the redshift errors is best modelled by a Gaussian-convolved Laplace
distribution. We adopt two different redshift error models for the luminous
and faint galaxies (see details in Appendix~\ref{app:zerr}). The typical
1$\sigma$ redshift error is about 10-15 ${\rm km\, s^{-1}}$. The redshift
errors (following the Gaussian-convolved Laplace distribution) are built into
our HOD model.

Following G15, we apply a Markov Chain Monte Carlo method to explore the HOD
parameter space. The likelihood $\propto \exp(-\chi^2/2)$ for a given set of
HOD parameters is determined by the $\chi^2$, contributed by the projected
2PCF $w_p(r_p)$, the redshift-space multipoles $\xi_0(s)$, $\xi_2(s)$ and
$\xi_4(s)$, and the observed galaxy number density $n_g$,
\begin{equation}
\chi^2= \bmath{(\xi-\xi^*)^T C^{-1} (\xi-\xi^*)}
       +\frac{(n_g-n_g^*)^2}{\sigma_{n_g}^2}, \label{eq:chi2}
\end{equation}
where $\bmath{C}$ is the full error covariance matrix and the data vector
$\bmath{\xi} = [\bmath{w_p},\bmath{\xi_0},\bmath{\xi_2},\bmath{\xi_4}]$. The
quantity with (without) a superscript `$*$' is the one from the measurement
(model). The covariance matrix is determined from 400 jackknife samples as
mentioned above \citep{Zehavi11,Guo13}. We apply a mean correction for the
bias effect in inverting the covariance matrix, as described in
\cite{Hartlap07} \citep[see also][]{Percival14}. We also apply a volume
correction of $1+V/V_{\rm sim}$ to the covariance matrix to account for the
model uncertainty caused by the the finite volume ($V_{\rm sim}=1\,h^{-3}{\rm
Gpc}^3$) of the MDPL simulation (G15). The error $\sigma_{n_g}$ on the number
density is determined from the variation of $n_g$ in the different jackknife
samples. The volume $V$ and mean number density $n_g$ of each luminosity
threshold sample are listed in Table~\ref{tab:sample}.

\section{Results}\label{sec:results}
\begin{figure*}
\includegraphics[width=0.8\textwidth]{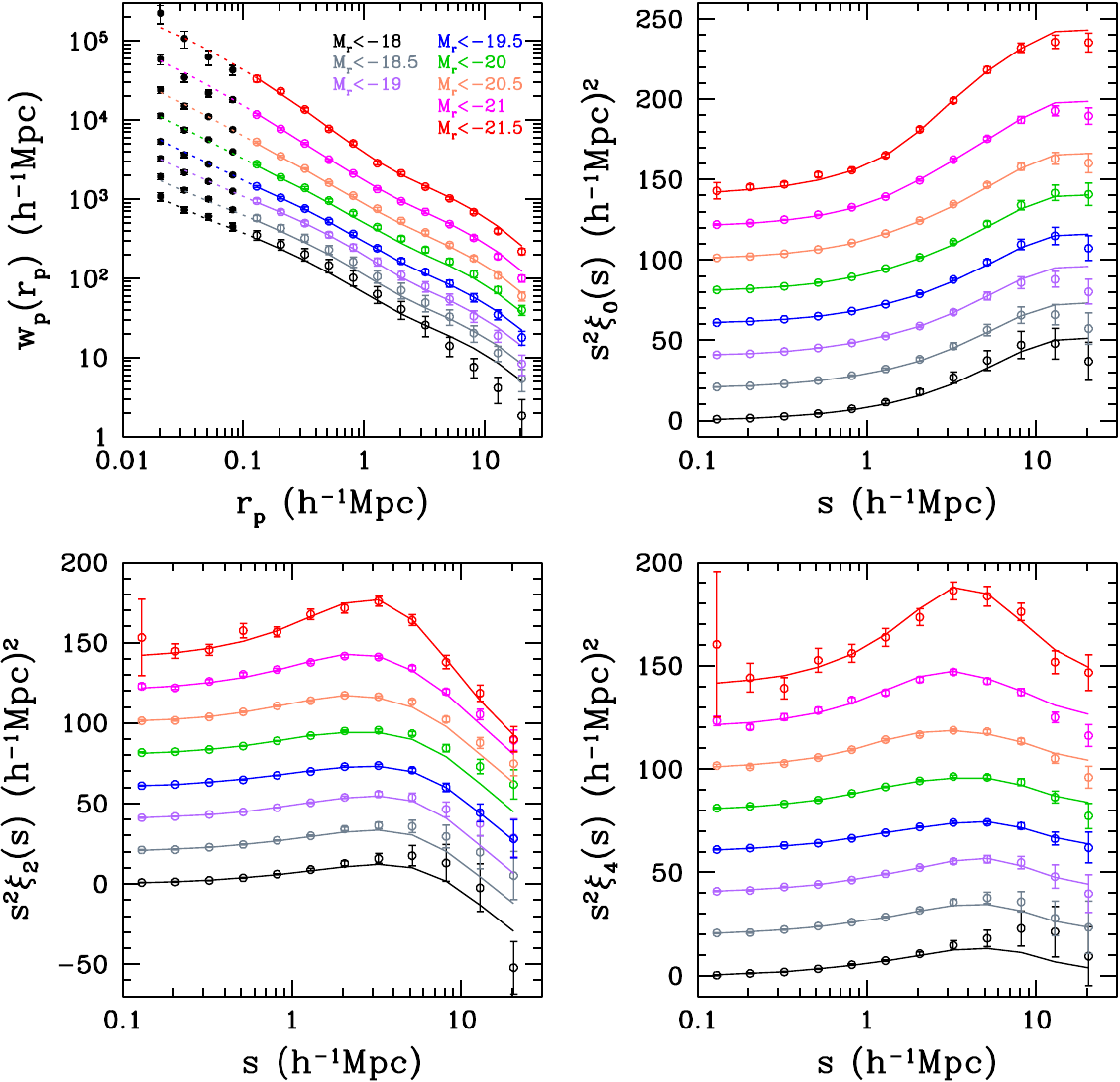}
\caption{HOD fittings to the four sets of 2PCF measurements in
Fig.~\ref{fig:wx}. The measurements from the data are shown by open circles,
while the HOD model fits are displayed as lines. For clarity, the
measurements of different luminosity threshold samples are separated by
$0.2{\rm dex}$ (for $w_p$) or $20 (\mpchi)^2$ (for $s^2\xi_{0,2,4}$) per
sample, starting from the $M_r<-18$ sample. In the top left panel of
$w_p(r_p)$, the measurements at $r_p<0.1\mpchi$ (below the fibre-collision
scale) are also shown, which are not included in the HOD fittings, while the
dashed lines are the predictions from the best-fitting HOD models.}
\label{fig:wxfit}
\end{figure*}

\subsection{Fitting Results and the Mean Occupation Function}

Fig.~\ref{fig:wxfit} shows the measurements and best-fitting HOD models for
the four sets of 2PCFs as in Fig.~\ref{fig:wx}. For clarify, offsets are
applied to both the data points and bestfitting curves, $0.2{\rm dex}$ for
$w_p$ and $20 (\mpchi)^2$ for $s^2\xi_{0,2,4}$ for each galaxy sample. As is
evident, our HOD model leads to remarkably good fits to all the luminosity
threshold samples, for both the projected and redshift-space 2PCFs. We choose
to fit 2PCFs on scales above $0.1\mpchi$ to reduce any possible systematic
effect in the fibre-collision correction (which is small according to
\citealt{Guo12}). With our bestfitting HOD model, we can predict the 2PCFs on
smaller scales. The dotted curves and the filled circles in the top left
panel of Fig.~\ref{fig:wxfit} show the prediction and the measurements on
scales below $0.1\mpchi$. The best-fitting models reproduce well the
measurements for all luminosity threshold samples down to $r_p=0.02\mpchi$,
suggesting both an accurate HOD model and a robust fibre-collision
correction.

\begin{table*}
\caption{Best-fitting HOD parameters and derived parameters for the
luminosity threshold samples} \label{tab:hod} \tabcolsep=0.11cm
\begin{tabular}{lrrrrrrrr}
\hline
$M_r^{\rm max}$ & -18.0 & -18.5 & -19.0 & -19.5 & -20.0 & -20.5 & -21.0 & -21.5\\
\hline
$\chi^2/{\rm dof}$ & $31.02/42$ & $34.77/42$ & $31.52/42$ & $32.29/42$ & $32.17/42$ & $46.24/42$ & $44.38/42$ &
$39.16/42$ \\
$\log M_{\rm min}$ & $11.18\pm0.14$ & $11.38\pm0.10$ & $11.58\pm0.09$  & $11.67\pm0.07$ & $11.95\pm0.06$ &
$12.23\pm0.04$ & $12.78\pm0.11$  & $13.53\pm0.10$  \\
$\sigma_{\log M}$ & $0.09\pm0.49$ & $0.23\pm0.22$ & $0.00\pm0.21$ & $0.01\pm0.21$  & $0.16\pm0.17$ & $0.18\pm0.13$ &
$0.49\pm0.13$ & $0.72\pm0.08$   \\
$\log M_0$ & $11.57\pm0.24$ & $11.73\pm0.16$ & $11.61\pm0.22$ & $11.80\pm0.13$ & $12.10\pm0.10$ & $12.42\pm0.12$ &
$12.59\pm1.59$ & $13.13\pm2.88$  \\
$\log M_1^\prime$ & $12.48\pm0.12$ & $12.71\pm0.10$ & $13.04\pm0.08$  & $13.07\pm0.06$ & $13.33\pm0.06$ &
$13.57\pm0.05$
& $13.99\pm0.07$ & $14.52\pm0.06$  \\
$\alpha$ & $0.97\pm0.07$ & $1.02\pm0.06$ & $1.12\pm0.04$ & $1.06\pm0.03$ & $1.08\pm0.03$ & $1.06\pm0.05$ &
$1.14\pm0.08$
& $1.14\pm0.16$ \\
$\alpha_c$ & $0.01\pm0.13$ & $0.01\pm0.09$ & $0.29\pm0.12$ & $0.28\pm0.07$ & $0.25\pm0.07$ & $0.29\pm0.04$ &
$0.27\pm0.04$ & $0.31\pm0.04$ \\
$\alpha_s$ & $0.95\pm0.05$ & $0.81\pm0.04$ & $0.77\pm0.03$ & $0.86\pm0.03$  & $0.84\pm0.03$ & $0.85\pm0.03$ &
$0.97\pm0.05$   & $1.05\pm0.08$   \\
\hline
$\log M_1$ & $12.53\pm0.10$ & $12.76\pm0.09$ & $13.06\pm0.07$ & $13.10\pm0.05$ & $13.35\pm0.05$  & $13.60\pm0.04$  &
$14.01\pm0.06$  & $14.55\pm0.06$  \\
$f_{\rm{sat}}$ & $26.87\pm1.53$  & $23.90\pm1.20$  & $20.74\pm1.03$  & $20.45\pm0.80$ & $17.87\pm0.75$  &
$15.67\pm0.57$
& $12.46\pm0.90$  & $7.74\pm0.74$   \\
$\sigma_c$ & $0.08\pm8.39$ & $0.37\pm6.15$ & $22.36\pm9.81$  & $22.85\pm4.17$ & $25.21\pm7.17$ & $35.05\pm5.13$  &
$50.24\pm7.95$  & $99.20\pm15.19$ \\
$\sigma_s$ & $146.24\pm13.68$ & $146.78\pm12.80$ & $173.85\pm11.53$& $197.32\pm9.01$& $234.58\pm11.83$&
$285.16\pm13.91$& $442.16\pm31.99$& $705.10\pm70.42$\\
\hline
\end{tabular}

\medskip
The halo mass is in units of $\msun$. The best-fitting $\chi^2$ per degrees
of freedom (dof) of the HOD modelling is also given. The degree of freedom of
each sample is calculated as ${\rm dof}=N_{\rm 2PCF}+1-N_{\rm par}$, where
the total number of data points ($N_{\rm 2PCF}+1$) is 49 (12 for each of
$w_p$ and $\xi_{0,2,4}$, plus one number density constraint), and $N_{\rm
par}=7$ is the number of HOD parameters. The derived parameter $\log M_1$ is
the mass of a halo that on average hosts one satellite galaxy. The satellite
fraction $f_{\rm{sat}}$ is in units of $\rm{per\,cent}$. The $\sigma_c$ and
$\sigma_s$ (in units of $\kms$) are the typical velocity dispersion of
central and satellite galaxies, respectively (see details in text).
\end{table*}

\begin{figure*}
\includegraphics[width=1.0\textwidth]{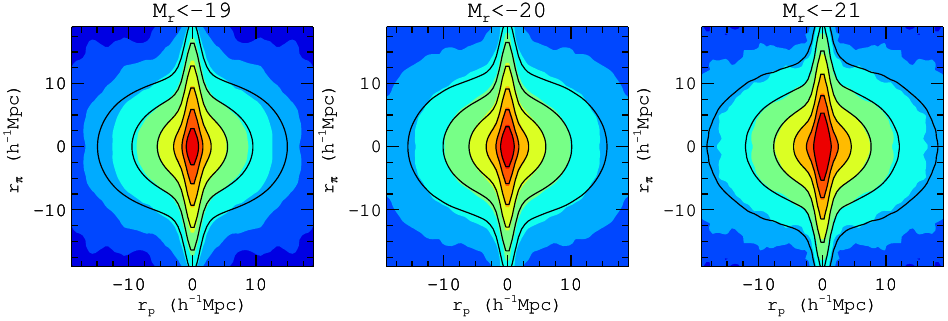}
\caption{ Comparisons between the measured redshift-space 3D 2PCF
$\xi(r_p,r_\pi)$ (color scales) and the prediction from the best-fitting HOD
model (black solid curves), for three representative luminosity-threshold
samples. The data and models have the same contour levels of 0.3, 0.5, 1,2,
5, 10, and 20. } \label{fig:xiplinear}
\end{figure*}
\begin{figure*}
\includegraphics[width=0.8\textwidth]{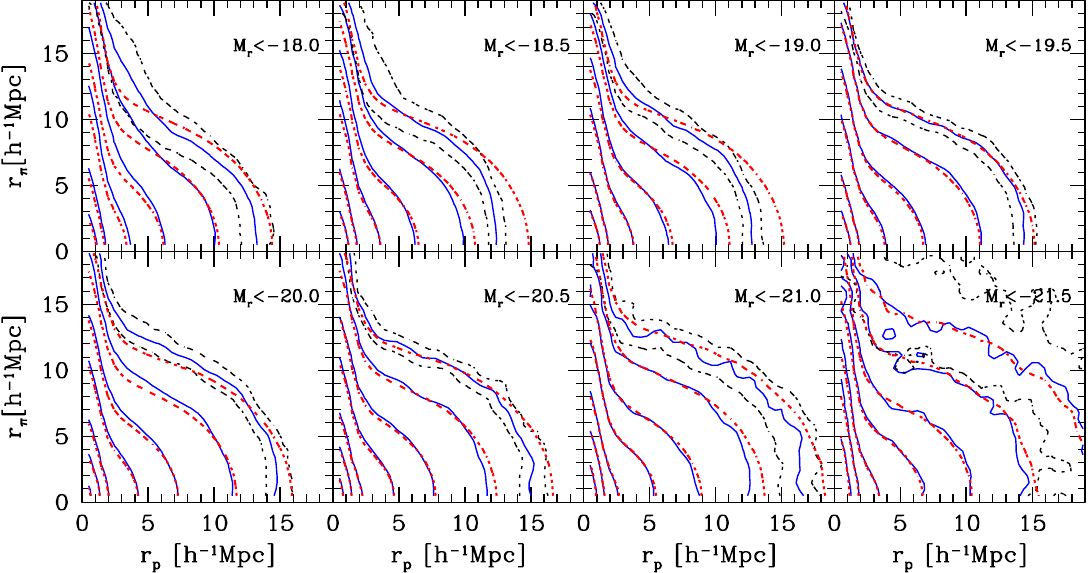}
\caption{Comparisons between the predicted 3D 2PCF $\xi(r_p,r_\pi)$ from the
best-fitting HOD models (red dotted curves) and the measurements (blue solid
curves) for different luminosity-threshold samples. Contour levels are set to
$\xi(r_p,r_\pi)=[0.3,0.5,1,2,5,10,20]$. The black dotted curves in each panel
are contours at the level of $\xi(r_p,r_\pi)=0.3$ with $\xi(r_p,r_\pi)$
shifting by $\pm 1\sigma$ measurement error. The measurements of
$\xi(r_p,r_\pi)$ in each panel (as well as the model curves) are smoothed
with a Gaussian kernel to reduce the noise on large scales.} \label{fig:xip}
\end{figure*}
The best-fitting HOD parameters are listed in Table~\ref{tab:hod} for the
different luminosity threshold samples. The $\chi^2/\rm{dof}$ of the fittings
confirm the adequacy of the model in fitting the data. The derived parameters
$M_1$ and $f_{\rm sat}$ are also displayed, together with the characteristic
central and satellite galaxy velocity dispersions
$\sigma_c=\alpha_c\sigma_v(M_{\rm min})$ and $\sigma_s=\alpha_s\sigma_v(M_1)$
(see \S~\ref{sec:vbias} for more details).

In our modelling, we choose to fit the projected 2PCF and the redshift-space
2PCF multipole $\xi_{0,2,4}$, not the 3D redshift-space 2PCF $\xi(r_p,r_\pi)$
directly. The reason is the large dimension of $\xi(r_p,r_\pi)$ (e.g. 240
data points per sample for 12 $r_p$ bins and 20 $r_\pi$ bins), which makes it
difficult to estimate a robust covariance matrix. However, with the
best-fitting models, we can predict $\xi(r_p,r_\pi)$ and compare to the
measurements as a cross-check. Such a comparison is shown in
Fig.~\ref{fig:xiplinear} for three representative luminosity-threshold
samples of $M_r<-19$, $-20$ and $-21$. Two main RSD effects show up in
$\xi(r_p,r_\pi)$. On large scales, the galaxy infall towards overdense
regions as well as the streaming of galaxies out of underdense regions
compresses the contours along the LOS direction, known as the Kaiser
squashing effect \citep{Kaiser87,Hamilton92}. On small scales, the random
motions of galaxies in virialized structures cause the $\xi(r_p,r_\pi)$
contours to appear stretched along the LOS direction, causing the FOG effect
\citep{Jackson72,Huchra88}. The best-fitting HOD models reproduce the two
features and the overall $\xi(r_p,r_\pi)$ measurements very well. In
particular, the agreement between the measured and predicted 3D 2PCFs on
small scales is remarkably good.

On large scales, there appears to be slight deviations of the predictions
from the measurements. which are not significant, given the measurement
errors. To see this and to have a comparison for all samples, in
Fig.~\ref{fig:xip} we compare the 3D 2PCF contours for both the best-fitting
predictions (red dotted curves) and the meausurements (blue solid curves). To
illustrate the uncertainties on the measurements of $\xi(r_p,r_\pi)$, we show
in each panel with black dotted curves the contours from $\pm 1\sigma$ of the
$\xi(r_p,r_\pi)$ measurements for the outmost level ($\xi(r_p,r_\pi)=0.3$).
The model fits of the $M_r<-18.5$ and $M_r<-19$ samples seem to be out of the
1$\sigma$ range of the measurements on large scales of $r_p>10\mpchi$, which
is consistent with the over-prediction of the model on large scales, as seen
in $w_p$ and $\xi_0$ in Fig.~\ref{fig:wxfit}. However, such deviations are
not significant given the highly correlated covariance matrix elements on
these scales. The largest difference in the FoG feature is seen in the
$M_r<-18.0$ sample, which is in fact not significant given the large error
bars in the measurement for this sample. Overall the model successfully
reproduces the luminosity dependent $\xi(r_p,r_\pi)$ measurements. As will be
discussed in \S~\ref{sec:vbias}, velocity bias is needed for the model to fit
the redshift-space clustering.

\begin{figure}
\includegraphics[width=0.43\textwidth]{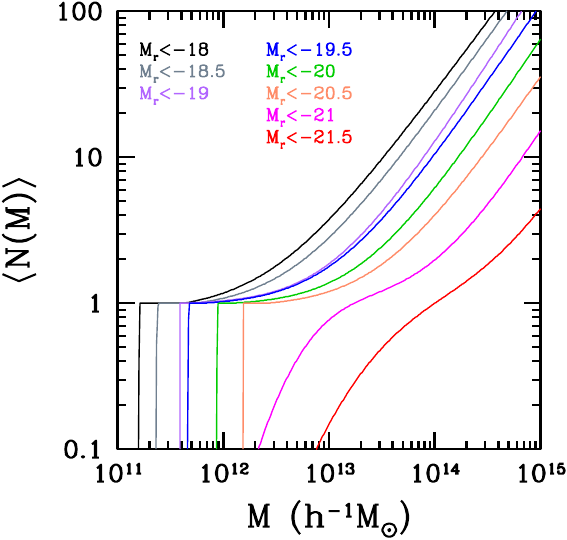}
\caption{Mean halo occupation functions of the best-fitting models for different luminosity threshold samples.
} \label{fig:hod}
\end{figure}
The mean halo occupation functions for the different luminosity threshold
galaxy samples are presented in Fig.~\ref{fig:hod}. The most significant
trend is that the host halo mass scale increases with the galaxy luminosity,
as expected from HOD modelling of projected 2PCF $w_p(r_p)$ from the same
SDSS data by \cite{Zehavi11} (see also \citealt{Zehavi05,Zheng07}). Compared
to the HOD model used in \cite{Zehavi11}, our model in this paper adopts
different cosmological parameters and halo definition. Furthermore, we shift
to a simulation-based model, rather than an analytic model. Finally, we
jointly fit the projected 2PCF $w_p$ and the redshift-space 2PCFs
$\xi_{0,2,4}$, while \cite{Zehavi11} only fit $w_p$. Accounting for these
differences, our results are in good agreement with those in \cite{Zehavi11}.
We note that the uncertainties in many HOD parameters (and the derived
satellite fraction) from the modelling in this paper appear to be larger than
those in \cite{Zehavi11} from modelling $w_p$ only. This can be attributed to
the differences in the models. The accuracy of the measured small-scale data
points exceed that of the analytic HOD model used in \cite{Zehavi11}, which
may be the reason of their large $\chi^2/$dof (2--3 for some cases). As a
consequence, the uncertainties in the HOD parameters can be artificially
underestimated. The simulation-based model used in this paper is a more
accurate model, leading to good values of $\chi^2/$dof and improved error
estimates in the parameters.

\begin{figure*}
\includegraphics[width=0.8\textwidth]{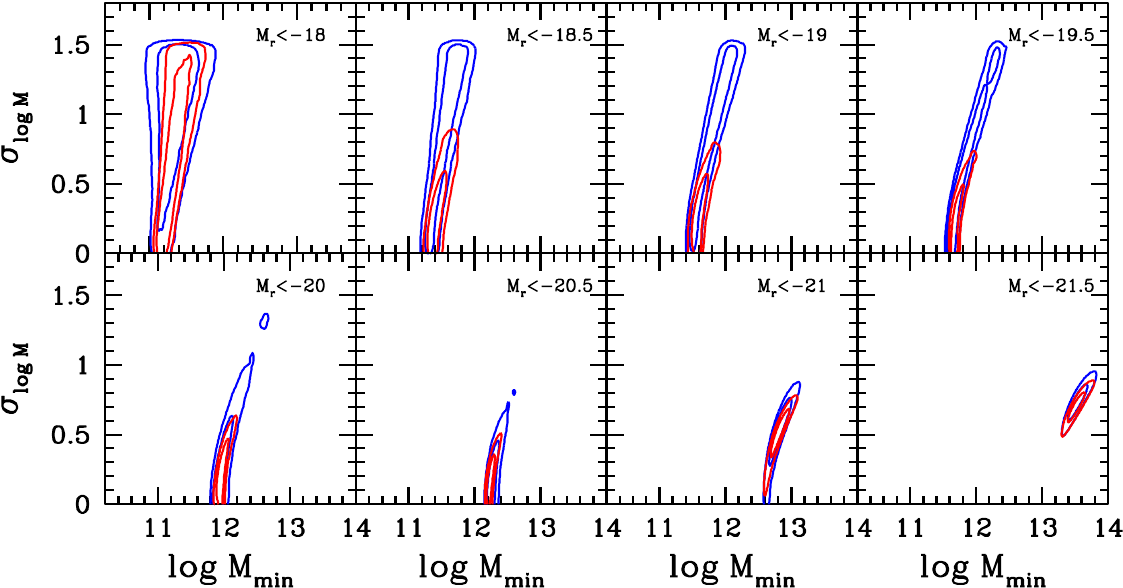}
\caption{Comparisons between the constraints on the HOD parameters $M_{\rm
min}$ and $\sigma_{\log M}$ from fitting $w_p$ (blue contours) and jointly
fitting $w_p$ and $\xi_{0,2,4}$ (red contours).} \label{fig:sigmin}
\end{figure*}
We also perform $w_p$-only fit with the simulation-based model and compare to
the results from fitting both $w_p$ and $\xi_{0,2,4}$. We find that
redshift-space 2PCFs help tighten the constraints on the HOD parameters. As
an example, we show in Fig.~\ref{fig:sigmin} the comparison of the
constraints on $M_{\rm min}$ and $\sigma_{\log M}$ from fitting $w_p$ only
(blue contours) and jointly fitting $w_p$ and $\xi_{0,2,4}$ (red contours).
We set a prior of $\sigma_{\log M}<1.5$ when fitting the data to have a
reasonable value of the scatter. Clearly, a substantial improvement with the
redshift-space 2PCFs is to narrow down the range of $\sigma_{\log M}$,
especially for less luminous samples (with the $M_r<-18$ as an exception,
which has a tighter $M_{\rm min}$).

Even though redshift-space 2PCFs help tighten the constraints on
$\sigma_{\log M}$, we note that for faint galaxy samples the cutoff profile
in the mean central occupation function is still not well constrained, as
indicated by the large errors (Table~\ref{tab:hod} and
Fig.~\ref{fig:sigmin}). It is consistent with a sharp cutoff at $M_{\rm
min}$, and in Fig.~\ref{fig:hod} we choose to plot the best-fitting models
with $\sigma_{\log M}\simeq 0$ for these samples. The constraints on $M_{\rm
min}$ and $\sigma_{\log M}$ mainly come from the galaxy bias (large scale
2PCF amplitude) and the galaxy number density. The galaxy bias is mainly
determined by haloes around $M_{\rm min}$. For faint samples, $M_{\rm min}$
is in the range that halo bias is insensitive to halo mass. As a consequence,
the galaxy bias is insensitive to the way of populating galaxies into haloes
of different masses around $M_{\rm min}$, i.e. insensitive to the change in
$\sigma_{\log M}$. A change in $\sigma_{\log M}$ can be easily compensated by
a slight change in $M_{\rm min}$ to maintain the galaxy number density.
Therefore, the cutoff profiles for faint samples are not well constrained.
Conversely, $\sigma_{\log M}$ is much better constrained for the luminous
samples as a result of the steep dependence of halo bias and halo mass
function on halo mass toward the high-mass end.

\begin{figure}
\includegraphics[width=0.4\textwidth]{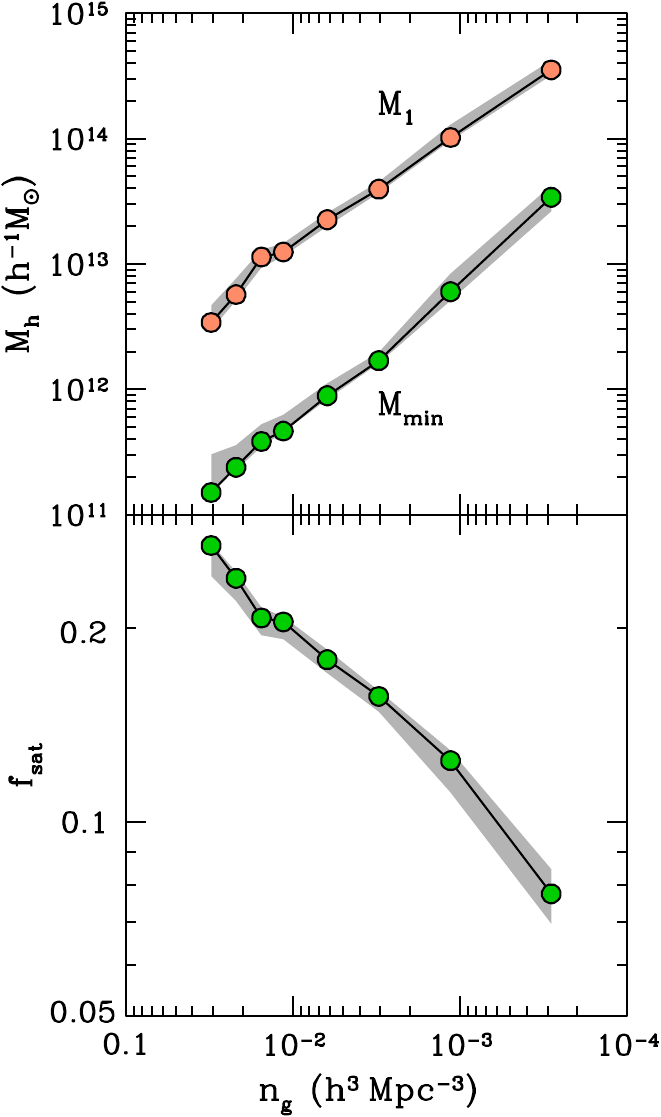}
\caption{Top: characteristic mass scales ($M_{\rm min}$ and $M_1$) of haloes hosting central galaxies and satellites as
a function of the sample number density $n_g$. Bottom: satellite fraction as a function of the sample number density.
The shaded area around each curve shows the $1\sigma$ uncertainties in the parameter.} \label{fig:m1min}
\end{figure}

Fig.~\ref{fig:m1min} shows the dependence of the characteristic mass scales
($M_{\rm min}$ for central galaxies and $M_1$ for satellite galaxies) and the
satellite fraction $f_{\rm sat}$ on the sample number density $n_g$. The
dependence of any of those parameters on $n_g$ roughly follows a power-law
form. As pointed out in \cite{Guo14}, the dependence of $M_{\rm min}$ on the
number density largely comes from the nearly power-law form of the halo mass
function over a large mass range. The mass $M_{\rm min}$ is mostly determined
by matching the halo number density with the galaxy number density, modulated
by $\sigma_{\log M}$. There is a trend that the ratio $M_1/M_{\rm min}$
decreases as the sample number density decreases (or the sample luminosity
increases), consistent with those found in literature
\citep[e.g.][]{Zehavi05,Skibba07,Zheng09,Zehavi11,Guo14,McCracken15,Skibba15}.
This is a manifestation of the halo mass dependent competition between
accretion of galaxies into haloes and destruction of galaxies inside haloes.
From the bottom panel of Fig.~\ref{fig:m1min}, we see that fainter galaxies
are more likely to be satellite galaxies in massive haloes. The satellite
fraction follows $f_{\rm sat}\simeq 0.1[\bar{n}_g/(10^{-3}h^3{\rm
Mpc}^{-3})]^{1/3}$, as also shown in other surveys \citep{Guo14}, which can
be used to estimate the satellite fraction given the number density of a
threshold galaxy sample. Note that the values of satellite fraction are
slightly lower than those inferred in \citet{Zehavi11}, which can be mostly
attributed to the difference in the definitions (hence the sizes) of haloes.

\subsection{Galaxy Velocity Bias}
\label{sec:vbias} 
\begin{figure*}
\includegraphics[width=0.8\textwidth]{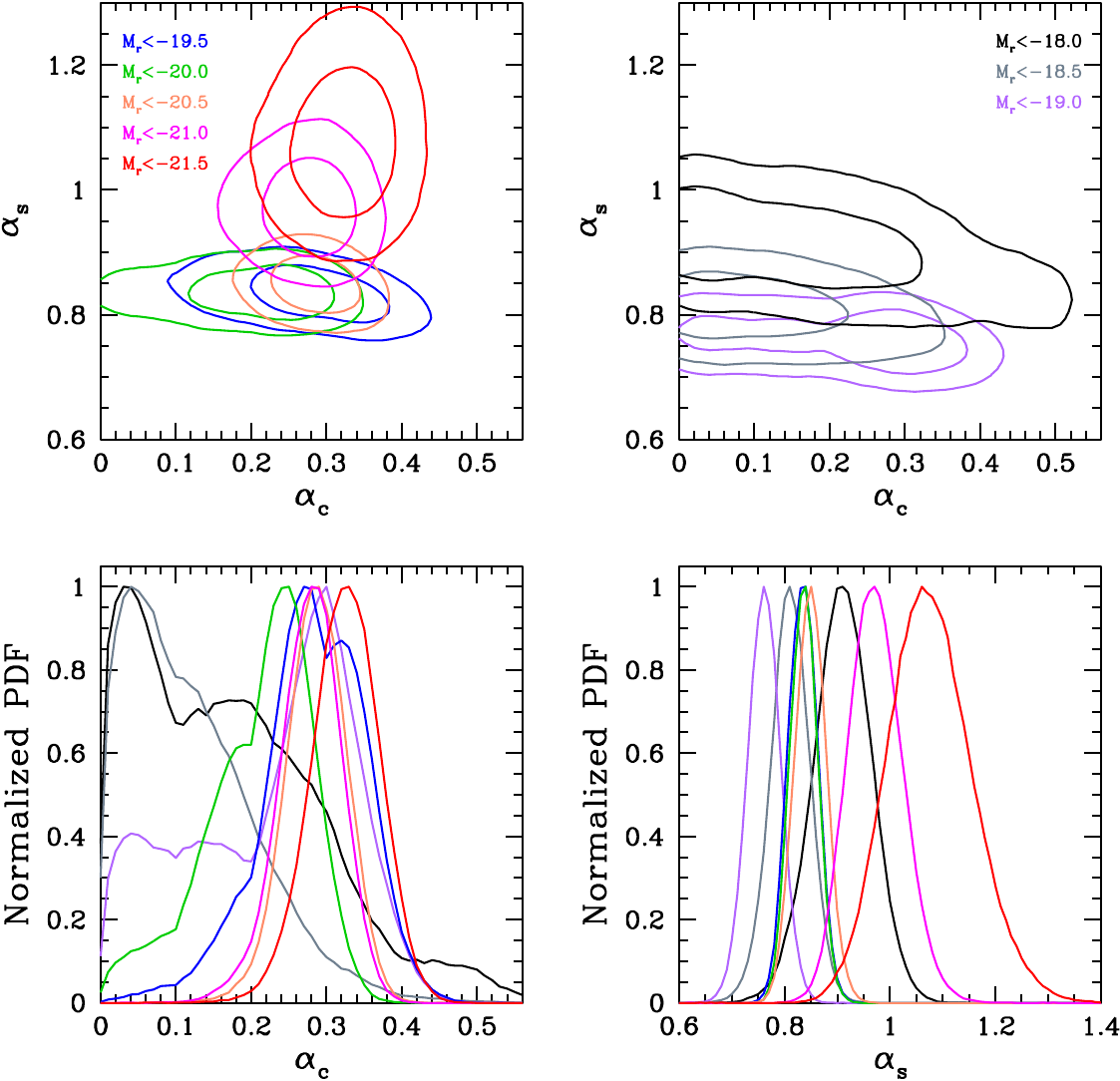}
\caption{Top panels: marginalized probability distributions of central and
satellite galaxy velocity bias parameters for the luminous (left) and faint
(right) galaxy samples. The contours show the 68 and 95 per cent confidence
levels for the two parameters. Bottom panels: 1D probability distribution of
the central (left)  and satellite (right) galaxy velocity bias parameters for
various luminosity-threshold samples.} \label{fig:alphacs}
\end{figure*}

The constraints on the galaxy velocity bias parameters are
shown in Fig.~\ref{fig:alphacs}, including the 68 per cent and 98 per cent
contours in the $\alpha_c$--$\alpha_s$ plane and the marginalized
distributions of $\alpha_c$ and $\alpha_s$ for each sample. In terms of the
tightness in the central galaxy velocity bias $\alpha_c$ constraints, the
luminous and faint samples show a dichotomy.

For the luminous samples (more luminous than $M_r=-19.5$), both the central
and satellite velocity bias parameters are well constrained, as shown in the
top left panel. The case without any galaxy velocity bias (i.e. $\alpha_c=0$
and $\alpha_s=1$) is far beyond the 95 per cent contours of all luminous
samples. That is, galaxy velocity bias is required to reproduce the
redshift-space clustering in the local universe for luminous samples.

The central velocity bias parameter $\alpha_c$ for luminous samples is about
0.3 (top-left and bottom-left panels in Fig.~\ref{fig:alphacs}). It shows no
significant dependence on galaxy luminosity. The existence of the central
velocity bias implies that these luminous central galaxies are not at rest at
the halo centres with respect to the bulk motion of the haloes. This reflects
the mutual (non)relaxation status of central galaxies and host haloes, which
are related to the merger history of the galaxies and the formation history
of haloes (see more detailed discussions in G15). The value of $\alpha_c$
inferred from our modelling of the redshift-space clustering is in agreement
with the estimates from galaxy group catalogues in the SDSS
\citep[e.g.][]{Bosch05}, which uses the mean velocity of satellites as a
proxy for the halo velocity.

The constraints on the central velocity bias parameter $\alpha_c$ are loose
for the three faint samples (with threshold luminosity fainter than
$M_r=-19.0$), as seen from the contours in the top-right panel and the
corresponding curves in the bottom-left panel of Fig.~\ref{fig:alphacs}. For
the $M_r<-19$ sample, $\alpha_c$ is different from zero only at the
2.5$\sigma$ level (see Table~\ref{tab:hod}). For the other two fainter
samples, $\alpha_c$ is consistent with zero. The loose constraints can be
partly attributed to the relatively large uncertainty in the clustering
measurements and in the jackknife covariance matrix estimate for the faint
samples, as the sample volumes are substantially smaller than those of the
luminous samples (see Fig.~\ref{fig:mlz} and Table~\ref{tab:sample}),
especially on large scales where $\alpha_c$ is mostly constrained (see Fig.6
of G15). The three faint samples have volumes that are 10\%, 24\%, and 43\%
that of the $M_r<-19.5$ sample, which has the smallest volume among the
luminous samples. The other possible cause of the loose constraints can be
the redshift errors. As the central velocity bias, in terms of velocity
dispersion (see below), approaches or drops below the level of redshift
errors (about 13 ${\rm km\, s^{-1}}$; see Appendix A), the sensitivity of RSD
to the central velocity bias is reduced, likely the case for the faint
samples.

For the satellite velocity bias, all samples show good constraints. Sample
volume becomes less important here, since the constraints mainly come from
the small-scale FoG effect (see Fig.6 of G15), where the uncertainties in the
measurements are small. However, the faintest sample ($M_r<-18$) may still be
affected by the small sample size and the noisy covariance matrix estimate,
and we should interpret the satellite velocity bias with caution. If we
neglect the $M_r<-18$ sample, the satellite velocity bias $\alpha_s$
constraints show a more or less continuous trend with luminosity, i.e. higher
values of $\alpha_s$ for more luminous sample. Probably more appropriately,
the satellite velocity bias constraints can be divided into two groups. For
the two most luminous samples (more luminous than $L_*$, corresponding to
$M_r=-20.44$; \citealt{Blanton03b}), $\alpha_s$ is consistent with unity.
That is, the motion of the luminous satellites closely follows that of the
dark matter. For samples with threshold luminosity fainter than $L_*$,
$\alpha_s$ is about 0.8 -- 0.85. That is, for those samples, satellites moves
more slowly than dark matter particles.

In a steady state, the spatial distribution and velocity distribution of
satellite galaxies inside haloes are related to each other. In our modelling,
we draw random dark matter particles for the position of satellites. That is,
we implicitly assumed that the spatial distribution of satellites follows
that of the dark matter, which is well described by the Navarro-Frenk-White
(NFW) profile \citep{Navarro97}. For the two luminous samples, $\alpha_s$ is
around unity, i.e., their satellite velocity distribution is consistent with
that of the dark matter. Therefore, based on the constraints we infer, an NFW
profile and associated velocity distribution for satellites are able to
explain the redshift-space clustering for the luminous sample. More modelling
efforts are needed to see whether other profiles and the corresponding
velocity distributions are preferred or not (see the tests in G15). For the
other, faint samples, the inferred $\alpha_s$ ($\sim$0.8 -- 0.85) differs
substantially from unity, inconsistent with the value for the NFW profile.
The result alone suggests that the spatial distribution of faint satellites
should deviate from the NFW profile.  Our current model, however, is not able
to provide more information on how significant a deviation it needs to be.
For an improved model, one can consider to parameterize the spatial profile
of satellites and solve for the corresponding velocity distribution in a
self-consistent manner, which is beyond the scope of this paper.

\citet{Watson10} and \cite{Watson12} analyzed the small-scale (down to
${\sim}0.01\mpchi$) clustering ($w_p$) of the SDSS Main galaxy sample and
luminous red galaxies. They found that the spatial distribution of faint
satellite galaxies (below $M_r=-20$) is consistent with the NFW profile,
while that of bright satellite galaxies deviates from the NFW profile (with a
steeper inner profile). These seem opposite to what we find from modelling
the redshift-space clustering. An improved model is necessary to constrain
the range of profiles allowed by the redshift-space clustering data and to
see whether this apparent difference is significant.

\begin{figure*}
\includegraphics[width=0.8\textwidth]{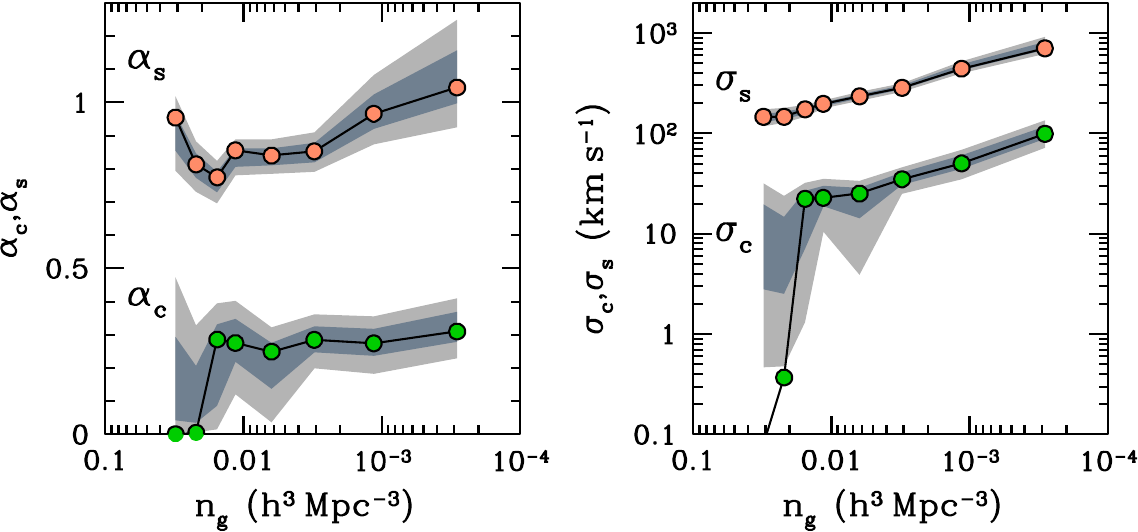}
\caption{Velocity bias and typical velocity dispersion. Left: central and
satellite galaxy velocity bias as a function of the sample number density
$n_g$. Right: dependence of the typical central and satellite galaxy velocity
dispersions, $\sigma_c$ and $\sigma_s$, on the sample number density. The
shaded area show the 68 per cent and 95 per cent confidence levels of the
parameters. The solid lines with circles are the best-fitting models. }
\label{fig:acng}
\end{figure*}

In the left panel of Fig.~\ref{fig:acng}, we summarize the constraints on the
velocity bias parameters $\alpha_c$ and $\alpha_s$ as a function of galaxy
number density (more luminous galaxy samples have lower number densities).
For galaxies $M_r<-19.5$ and brighter, there is no significant dependence of
$\alpha_c$ on number density. The three faint galaxy samples (with the
highest number densities) seem to have lower values of $\alpha_c$, consistent
with zero, but with large error bars. Since the host halo mass increases as
the galaxy luminosity increases (or as the number density decreases), the
trend also indicates the weak dependence of $\alpha_c$ on the host halo mass.
With the faintest sample excluded (smallest sample volume), the dependence of
satellite velocity bias $\alpha_s$ on the luminosity for faint galaxies is
also weak, while $\alpha_s$ shows a clear increase with luminosity for
$M_r<-21$.

To interpret the velocity bias results, the more meaningful physical
quantities are the velocity dispersions of central and satellite galaxies
inside haloes, denoted as $\sigma_c$ and $\sigma_s$, respectively. Given the
velocity bias parameters, velocity dispersions depend on halo mass, and we
choose to evaluate typical values in representative haloes.  For central
galaxies, the velocity bias constraint is mainly contributed from haloes
around $M_{\rm min}$, and we compute the typical central galaxy velocity
dispersion as $\sigma_c=\alpha_c \sigma_v(M_{\rm min})$. For satellite
galaxies, the effect of the velocity bias on clustering comes from haloes
around $M_1$, and the typical satellite velocity dispersion is computed as
$\sigma_s=\alpha_s\sigma_v(M_1)$. The right panel of Fig.~\ref{fig:acng}
shows the dependencies of these typical velocity dispersions on sample number
density, which roughly follow power-law relations,
$\sigma_c\simeq55\kms[\bar{n}_g/(10^{-3}h^3{\rm Mpc}^{-3})]^{-0.45}$ and
$\sigma_s\simeq437\kms[\bar{n}_g/(10^{-3}h^3{\rm Mpc}^{-3})]^{-1/3}$.

The existence of galaxy velocity bias reflects the dynamical evolution of
galaxies inside haloes. For example, infalling satellites experience tidal
striping and dynamical friction in the dark matter haloes, affecting its
velocity distribution. For central galaxies, the existence of velocity bias
indicates that central galaxy and the host haloes are not mutually relaxed. A
likely cause can be the halo mergers and the subsequent galaxy mergers. Our
results can be used to assess the dependence of the degree of relaxation
after mergers on halo mass (or galaxy number density). For mergers of haloes
of similar mass, the mean pairwise infall velocity $v_{12}$ on large scales
is proportional to the bias factor \citep{Sheth01,Zhang04}. With the
luminosity-dependent bias factor in \cite{Zehavi11}, we find that the bias
factor approximately scales as $n_g^{-0.11}$, which means that the pairwise
infall velocity $v_{12}\propto n_g^{-0.11}$ before merger. The central
velocity dispersion constrained from the RSD ($\alpha_c\propto n_g^{-0.45}$)
is much steeper than the infall velocity. We therefore conclude that central
galaxies in lower mass haloes are more relaxed with respect to the host
haloes, compared with their counterparts in more massive haloes, consistent
with an overall earlier formation and thus more time for relaxation of the
lower mass haloes.

\subsection{Dependence of Velocity Bias on Cosmology ($\Omega_m$)}

\begin{figure}
\includegraphics[width=0.42\textwidth]{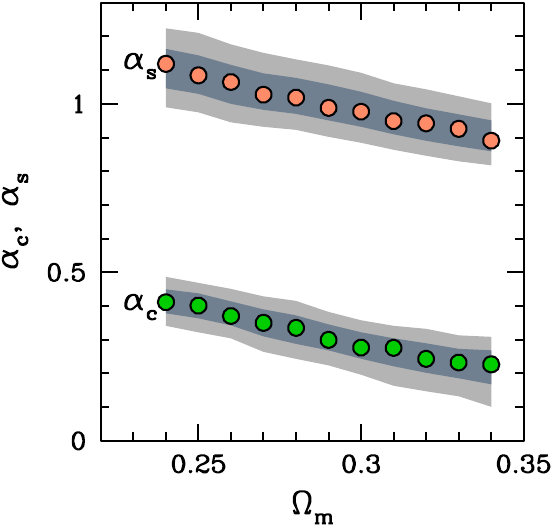}
\caption{Dependence of central and satellite galaxy velocity bias on the cosmological parameter $\Omega_m$. The central
and satellite velocity bias are displayed in circles of different colours. The shaded area show the 68 and 95 per cent
confidence levels of the parameters.
} \label{fig:fsigma}
\end{figure}

The velocity bias constraints we infer rely on the MultiDark simulation with
the assumed cosmology. The cosmological parameters are close to the results
from Planck \citep{Planck14}. It is useful to see whether the velocity bias
constraints are robust against reasonable change in cosmology. Here, we limit
our investigation to the change in $\Omega_m$. Instead of using simulations
with different $\Omega_m$ parameters, we use the appropriate scaling
relations with the MDPL simulation for the corresponding change in the dark
matter halo properties when varying $\Omega_m$.

According to \cite{Zheng02} and \cite{Tinker06}, if two simulations have the
identical initial matter fluctuation spectrum (including amplitude and phase)
but different values of $\Omega_m$, there exists a simple relation between
simulation outputs at a given linear growth factor $G$. There are
correspondences between haloes in the two simulations, and the corresponding
haloes have the same radius with the mass scaling with $\Omega_m$. Halo
velocity scales with the growth rate $f\equiv d\ln G/d\ln a$, with $a$ the
scale factor. The internal velocity dispersion inside haloes scales with
$\Omega_m^{0.5}$. We therefore modify our default simulation output by
scaling the halo mass $M$, halo velocity $v_h$, dark matter particle velocity
$v_p$, and the 1D velocity dispersion $\sigma_v$ of dark matter particles in
haloes in the following way,
\begin{eqnarray}
M       &=& \left(\Omega_m/\Omega_{m,0}\right)M_0,\\
v_h     &=& \left(f/f_0\right) v_{h,0},\\
v_p-v_h &=& \left(\Omega_m/\Omega_{m,0}\right)^{1/2}(v_{p,0}-v_{h,0}),\\
\sigma_v&=& \left(\Omega_m/\Omega_{m,0}\right)^{1/2}\sigma_{v,0},
\end{eqnarray}
where the symbols with subscript `0' denote the values in the fiducial MDPL
simulation. The halo mass function also changes accordingly in each scaled
simulation. We build eleven scaled simulation catalogues, varying $\Omega_m$
from $0.24$ to $0.34$ with a step of $\Delta\Omega_m=0.01$. We then apply our
HOD model to the $M_r<-21$ sample based on the eight scaled simulations to
constrain the velocity bias parameters.

We show in Fig.~\ref{fig:fsigma} the dependence of the velocity bias
parameter constraints on $\Omega_m$. Both the central and satellite velocity
biases decrease with increasing $\Omega_m$, as expected. Increasing
$\Omega_m$ leads to higher halo-halo and internal halo velocity dispersions,
and the velocity bias parameters decrease to compensate such a change to
match the redshift-space clustering. For the range of $\Omega_m$ considered
here, both velocity bias parameters are well constrained. In particular, the
central velocity bias differs from zero for all the cases, which shows that
our constraints are robust against reasonable changes in cosmology.

We note that the luminosity-threshold samples used in this
paper are constructed over different redshift ranges according to the range
for which they are volume limited (see Fig.~\ref{fig:mlz} and
Table~\ref{tab:sample}). If we were to apply the cosmology change and compare
the results among different samples, the difference in the sample mean
redshifts needs to be accounted for, which leads to small effective cosmology
changes. As mentioned in \S~\ref{sec:simmod}, our test with the model built
on the $z=0.1$ simulation output shows that the inferred HOD parameters are
within the uncertainties of those from the default model built on the $z=0$
output. So the effect of cosmology change from the difference in mean
redshift is small. Furthermore, here we only focus on one sample and aim to
see the sensitivity of velocity bias constraints to cosmology, and the
effective cosmology change among different samples becomes irrelevant. With
one sample, there is another effect related to the cosmology change. In
principle, when varying the cosmological parameter $\Omega_m$, we need to
rescale or remeasure the 3D 2PCFs and then perform the modelling. Otherwise,
the Alcock-Paczynski effect \citep{Alcock79} would introduce an additional
distortion in the 2PCFs. However, since the galaxy samples are local, with
$z\sim 0$, the comoving distance is insensitive to $\Omega_m$ and the effect
is tiny. For example, at the mean redshift of the $M_r<-21$ sample, $z\sim
0.12$, the comoving distance changes by less than 1\% in the whole range of
$\Omega_m$ we study here, which has little effect on the 2PCF measurements.
We therefore do not rescale or remeasure the 2PCFs changing the cosmology.

Relevant to the RSD effect, a more interesting change in cosmology is in the
combination of the growth rate and fluctuation amplitude, $f\sigma_8$.
Because of the high precision measurements, the small-scale redshift-space
clustering can help tighten the constraints on $f\sigma_8$
\citep[e.g.][]{Reid14}, providing potentially stringent tests to the
$\Lambda$CDM cosmology and theory of gravity. We reserve such an
investigation on the $f\sigma_8$ constraints with the SDSS Main galaxies for
a future work.

\section{Summary and Discussions}\label{sec:conclusions}

We measure the projected and redshift-space 2PCFs for volume-limited,
luminosity-threshold samples of SDSS Main galaxies, on small to intermediate
scales ($0.1$--$25\mpchi$). The measurements are interpreted within the HOD
framework to infer the relation between galaxies and dark matter haloes. In
particular, the RSD effects in the redshift-space 2PCFs enable us to
constrain the kinematics of central and satellite galaxies inside dark matter
haloes and infer the difference between the motions of galaxies and dark
matter.

It is the first time that the redshift-space clustering of local galaxies is
accurately measured on scales as small as $\sim$$0.1\mpchi$. The measurements
become possible with the accurate fibre-collision correction method developed
in \cite*{Guo12}, which makes use of the resolved collided galaxy pairs in
tile overlap regions to recover the small scale clustering. Previous
measurements \citep[e.g.][]{Hawkins03,Zehavi05} rely on either angular or
nearest neighbour fibre-collision corrections, which results in systematics
at the level of the data precision \citep{Guo12}. With our measurements, we
find that both the projected and redshift-space 2PCFs show a clear dependence
on galaxy luminosity, generally with a higher clustering amplitude for more
luminous galaxies on both small and large scales. The dependence on
luminosity becomes stronger for galaxies above $\sim$$L_*$. The overall trend
is consistent with previous results based on the projected 2PCFs
\citep[e.g.][]{Zehavi11}.

To interpret the measurements, similar to G15, we resort to an accurate HOD
model based on a high-resolution $N$-body simulation. In addition to the mean
halo occupation function, the model also parameterizes the central and
satellite galaxy velocity bias. For the first time, a halo-based model is
applied to model the measured luminosity dependent small- and
intermediate-scale redshift-space clustering of local galaxies. Previous
studies usually focus on relatively large scales and adopt a streaming model
\citep[e.g.][]{Peacock01,Hawkins03,Bel14,Howlett15}. The commonly inferred
quantities include the linear redshift distortion parameter and the mean
pairwise velocity dispersion of galaxies \citep[e.g][]{Cabre09} or its scale
dependence \citep[e.g.][]{Li07}. Our halo-based model, as applied in G15 (see
\citealt{Reid14} for a similar model), makes use of the kinematic information
of haloes and parameterizes the galaxy velocity distribution on top of it,
rather than an overall mean velocity dispersion. It allows us to constrain
the occupation and kinematic distribution of galaxies at the level of dark
matter haloes, a more informative extraction from galaxy redshift-space
clustering data. We find that the model is able to successfully reproduce the
observed projected 2PCF $w_p(r_p)$, the redshift-space multipole moments
$\xi_0(s)$, $\xi_2(s)$ and $\xi_4(s)$, and the 3D 2PCF $\xi(r_p,r_\pi)$, on
all scales for all SDSS luminosity-threshold samples (Fig.~\ref{fig:wxfit}
and Fig.~\ref{fig:xip}).

Consistent with previous work that only model the projected 2PCF
\citep[e.g.][]{Zheng07,Zehavi11}, we find that the clustering trend with
luminosity can be explained by the fact that more luminous galaxies reside in
more massive haloes. The characteristic halo masses, $M_{\rm min}$ (for
central galaxies) and $M_1$ (for satellite galaxies), increase with
increasing luminosity threshold. The satellite fraction, $f_{\rm sat}$, drops
as the luminosity threshold increases. The dependence of $M_{\rm min}$,
$M_1$, and $f_{\rm sat}$ on the sample number density (which is directly
related to the sample luminosity threshold) can be well described by
power-law relations. Compared to the $w_p$-only modelling results, the
redshift-space 2PCFs help tighten the constraints on the HOD parameters.
However, we find that for the faint galaxy samples (with luminosity threshold
below $L_*$), the cutoff profile (characterized by the parameter
$\sigma_{\log M}$) in the mean occupation function of central galaxies is
still loosely constrained.

Besides the above results, the brand-new outcomes from our modelling are the
constraints on galaxy kinematics inside haloes, coming from the RSD effects
on both small and large scales. The redshift-space clustering data require
the existence of a non-zero central galaxy velocity bias of about $0.3$ for
luminous samples (with threshold luminosity $M_r<-19.5$), while for faint
samples the central velocity bias parameters are loosely constrained but
consistent with the above value. That is, in the rest-frame (centre-of-mass
frame) of a halo, the central galaxy on average moves at a speed about 30\%
that of dark matter particles. The central galaxy velocity bias in our model
is in agreement with the estimates from the galaxy group catalogues in the
SDSS \citep[e.g.][]{Bosch05}. This mutual non-relaxation between central
galaxies and dark matter haloes can result from mergers and dynamical
evolution of galaxies and haloes. Converting the value to physical speed and
comparing to the typical infall velocity before merging, our results imply
that galaxies in lower mass haloes are more relaxed with respect to the host
haloes, consistent with their earlier formation time. Further theoretical
investigation of such an evolution paradigm using high-resolution
cosmological hydrodynamic galaxy formation simulations will be pursued in
future work.

For satellite galaxies, we find that the two most luminous samples have
satellite velocity bias consistent with unity, which means that they follow
closely the motion of dark matter. The satellite velocity bias $\alpha_s\sim
0.85$ for fainter samples implies that fainter satellites move more slowly
than dark matter. If satellite motion is in a steady state, the result
suggests that the spatial distribution profile of faint satellites should
differ from that of dark matter. More informative constraints need an
improved model that allow for a self-consistent treatment of the spatial and
kinematic distributions of satellites, which would also lead to tighter
constraints than using only projected spatial clustering measurements
\citep[e.g.][]{Watson12,Wang14}. An improved model can also include the
effect of halo assembly bias to see its influence on the HOD (e.g.
\citealt{Zentner14}; but see \citealt{Lin15}) and velocity bias. The
existence of satellite galaxy velocity bias affects any dynamical inference
based on satellite velocity dispersions. For example, the effect needs to be
taken into account when using the velocity dispersion of the galaxy cluster
members to estimate the halo mass of galaxy clusters \citep{Goto03,Old13}.

In G15, velocity bias is inferred for a sample of $z\sim 0.5$ luminous
galaxies ($n_g=2.19\times 10^{-4}h^3{\rm Mpc}^{-3}$). Converted to the same
velocity bias definition used in this paper (i.e. in the centre-of-mass frame
of haloes), their result on the velocity dispersion for central galaxies in
haloes of $\sim 10^{13.35}\msun$ is about $96\pm 8 \,{\rm km\, s^{-1}}$. On
average, these haloes evolve to $\sim 10^{13.55}\msun$ haloes at z$\sim 0$
\citep[e.g.][]{Zhao09}, around $M_{\rm min}$ of the $M_r<-21.5$ sample
($n_g=2.86\times 10^{-4}h^3{\rm Mpc}^{-3}$). We thus can have an approximate
connection between the two samples at $z\sim 0.5$ and $z\sim 0$ from the halo
evolution, following the same spirit as in \citet{Zheng07}, which is also
supported by the number density comparison. The central galaxy velocity
dispersion for the $M_r<-21.5$ is about $99\pm 15 \, {\rm km\, s^{-1}}$. Such
a preliminary analysis shows that the velocity dispersion for luminous
central galaxies has not evolved much since $z\sim 0.5$ (in fact, it has
marginally increased, but only at a $\sim 0.3\sigma$ level). We can assume a
simple circular motion of the central galaxy in the inner NFW halo to study
the implication of the result. With the $z\sim 0.5$ velocity bias result, the
radius of the orbit can be inferred to be about 0.4\% of the virial radius of
the halo (G15). The corresponding dynamical friction time scale is estimated
to be $\sim$0.1 Myr, much shorter than the 3.7 Gyr time span from $z\sim 0.5$
to $z\sim 0.1$. Therefore, a substantial central velocity bias at $z\sim0$
indicates that these luminous central galaxies {\it and} their host haloes
(or the cores of haloes and the rest of the haloes) may have been constantly
disturbed by galaxy and halo mergers. We expect that velocity bias of
galaxies inferred from different redshifts can help study the dynamical
evolution of galaxies and haloes and test galaxy formation theory.

The galaxy peculiar velocity field is directly related to the growth of
structures in the universe. RSD effects can be used to constrain cosmology,
especially the growth rate (i.e. the $f\sigma_8$ parameter) to test the
theory of gravity \citep[e.g.][]{Guzzo08,Percival09}. The statistical power
of small-scale RSD measurements have the potential to greatly tighten the
constraints \citep[e.g.][]{Tinker06,Tinker07,Reid14}. The effect of velocity
bias needs to be taken into account when using small-scale redshift-space
clustering to constrain cosmology. We leave an investigation on cosmological
applications based on our measurements and model for a future work.

\section*{Acknowledgements}
We thank Cheng Li for useful discussions. We thank the anonymous referee for
helpful comments. This work is supported by the 973 Program (No.
2015CB857003). HG acknowledges the support of the 100 Talents Program of the
Chinese Academy of Sciences. ZZ was partially supported by NSF grant
AST-1208891 and NASA grant NNX14AC89G. IZ acknowledges support, during her
sabbatical in Durham, from STFC through grant ST/L00075X/1, from the European
Research Council through ERC Starting Grant DEGAS-259586 and from a CWRU
ACES+ ADVANCE Opportunity Grant. Support for PSB was provided by a Giacconi
Fellowship. CC, GF and FP acknowledge support from the Spanish MICINNs
Consolider-Ingenio 2010 Programme under grant MultiDark CSD2009-00064, MINECO
Centro de Excelencia Severo Ochoa Programme under grant SEV-2012-0249, and
MINECO grant AYA2014-60641-C2-1-P. GY acknowledges financial support from
MINECO (Spain) under research  grants AYA2012-31101 and FPA2012-34694.

We gratefully acknowledge the use of the High Performance Computing Resource
in the Core Facility for Advanced Research Computing at Case Western Reserve
University, the use of computing resources at Shanghai Astronomical
Observatory, and the support and resources from the Center for High
Performance Computing at the University of Utah. The MultiDark database was
developed in cooperation with the Spanish MultiDark Consolider Project
CSD2009-00064. The MultiDark-Planck (MDPL) simulation suite has been
performed in the Supermuc supercomputer at LRZ using time granted by PRACE.

Funding for the SDSS and SDSS-II has been provided by the Alfred P. Sloan
Foundation, the Participating Institutions, the National Science Foundation,
the U.S. Department of Energy, the National Aeronautics and Space
Administration, the Japanese Monbukagakusho, the Max Planck Society, and the
Higher Education Funding Council for England. The SDSS Web Site is
http://www.sdss.org/.

The SDSS is managed by the Astrophysical Research Consortium for the
Participating Institutions. The Participating Institutions are the American
Museum of Natural History, Astrophysical Institute Potsdam, University of
Basel, University of Cambridge, Case Western Reserve University, University
of Chicago, Drexel University, Fermilab, the Institute for Advanced Study,
the Japan Participation Group, Johns Hopkins University, the Joint Institute
for Nuclear Astrophysics, the Kavli Institute for Particle Astrophysics and
Cosmology, the Korean Scientist Group, the Chinese Academy of Sciences
(LAMOST), Los Alamos National Laboratory, the Max-Planck-Institute for
Astronomy (MPIA), the Max-Planck-Institute for Astrophysics (MPA), New Mexico
State University, Ohio State University, University of Pittsburgh, University
of Portsmouth, Princeton University, the United States Naval Observatory, and
the University of Washington.

\bibliographystyle{mnras}

\begin{appendix}

\section{The Redshift Error Distribution} \label{app:zerr}
\begin{figure*}
\includegraphics[width=0.6\textwidth]{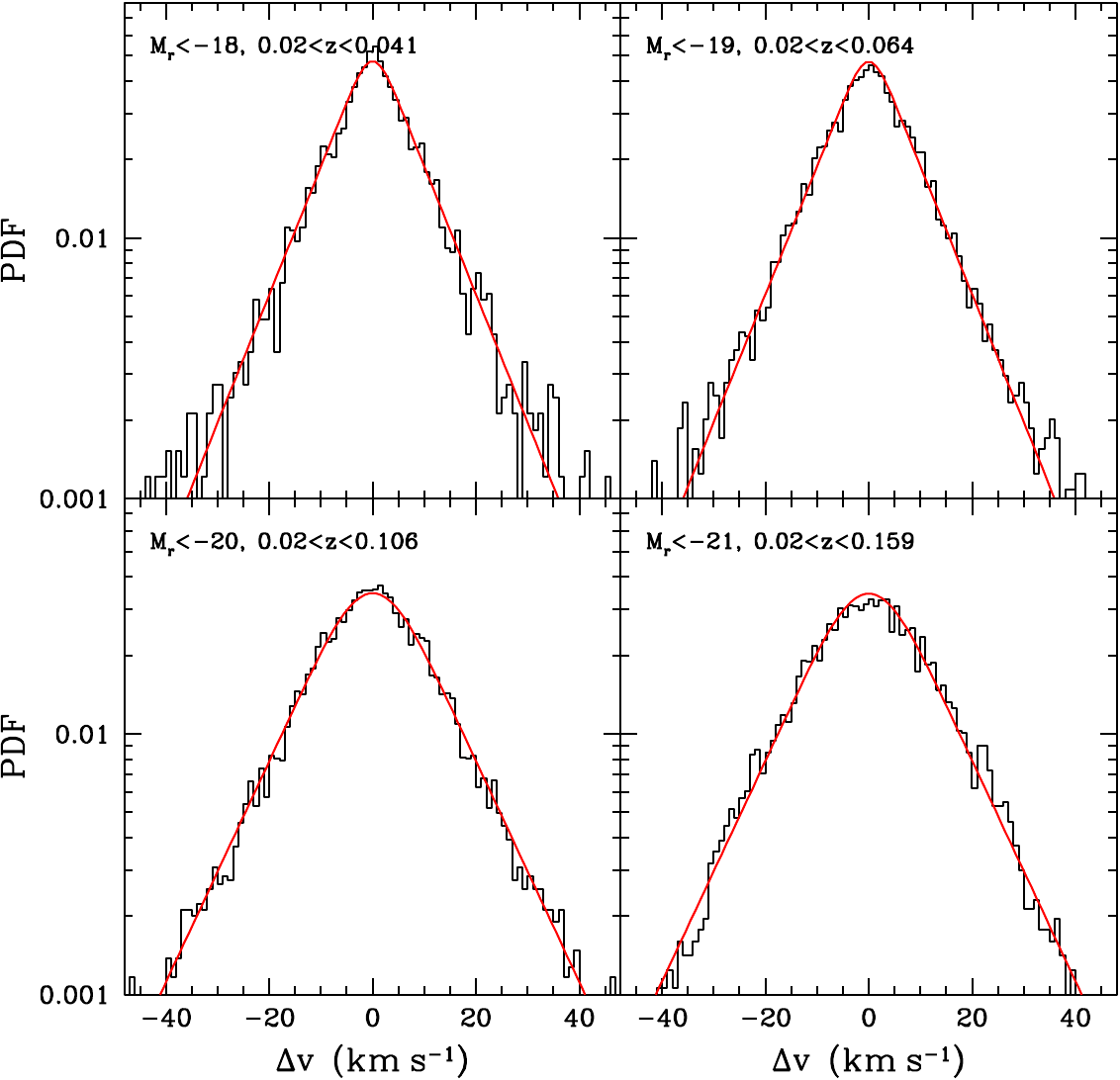}
\caption{ Redshift error distribution of SDSS Main galaxies with different
luminosity thresholds. In each panel, the histogram is obtained from galaxies
with repeat spectroscopic observations, and the red curve is the best-fitting
Gaussian-convolved Laplace distribution. See the text. } \label{fig:zerr}
\end{figure*}

In order to develop an accurate HOD model to apply to the redshift-space
clustering measurements in SDSS, we need to carefully account for the
distribution of redshift measurement errors of galaxies. The effect of
redshift errors is to add apparent peculiar velocity dispersion to galaxies,
which, if not accounted for, introduces apparent velocity bias component.
This is especially important for faint galaxies, because the potentially
small central and satellite velocities inside haloes make the constraints
more vulnerable to redshift errors.

To properly investigate the redshift errors in the SDSS Main galaxies, we
make use of all the galaxies with repeat spectra for each
luminosity-threshold sample to derive the distribution of redshift
measurement errors. In particular, for a galaxy with $n$ observations and
hence $n$ redshift measurements, we construct the estimator of the velocity
error in each measurement as
\begin{equation}
\Delta v_i = \sqrt{\frac{n}{n-1}}\frac{c(z_i-\bar{z})}{1+\bar{z}},
\end{equation}
where $z_i$ is the $i$--th redshift measurement for this galaxy ($i=1, ...,
n$) and $\bar{z}$ is the mean of the $n$ measurements, and the factor
$\sqrt{n/(n-1)}$ makes the estimator unbiased.

The histograms in Fig.~\ref{fig:zerr} show the probability distribution of
the redshift errors in terms of the peculiar velocity errors $\Delta v$ in
four different luminosity-threshold samples. We find that the distribution
has more extended tails than a naively assumed Gaussian distribution. The
extended part follows closely the double exponential distribution or the
Laplace distribution (appearing as straight lines in the plot), while the
central part can be described by a Gaussian core (less sharply peaked than
the Laplace distribution). In fact, the distribution can be remarkably fitted
by a Gaussian-convolved Laplace distribution, shown as the red curves of
Fig.~\ref{fig:zerr}. A random deviate $\Delta v$ for such a distribution can
be obtained by the sum of two independent Gaussian and Laplace random
numbers, i.e. $\Delta v=\Delta v_{\rm gau}+\Delta v_{\rm exp}$, with the
distribution functions of $\Delta v_{\rm gau}$ and $\Delta v_{\rm exp}$ as
\begin{eqnarray}
f_{\rm gau}(\Delta v_{\rm gau})&=&\frac{1}{\sqrt{2\pi}\sigma_{\rm gau}}\exp\left(-\frac{\Delta v_{\rm
gau}^2}{2\sigma_{\rm gau}^2}\right),\\
f_{\rm exp}(\Delta v_{\rm exp})&=&\frac{1}{\sqrt{2}\sigma_{\rm exp}}\exp\left(-\frac{\sqrt{2}|\Delta v_{\rm
exp}|}{\sigma_{\rm exp}}\right).
\end{eqnarray}
The parameters $\sigma_{\rm gau}$ and $\sigma_{\rm exp}$ are the standard
deviations for the Gaussian and Laplace distributions, respectively.

\begin{figure}
\includegraphics[width=0.4\textwidth]{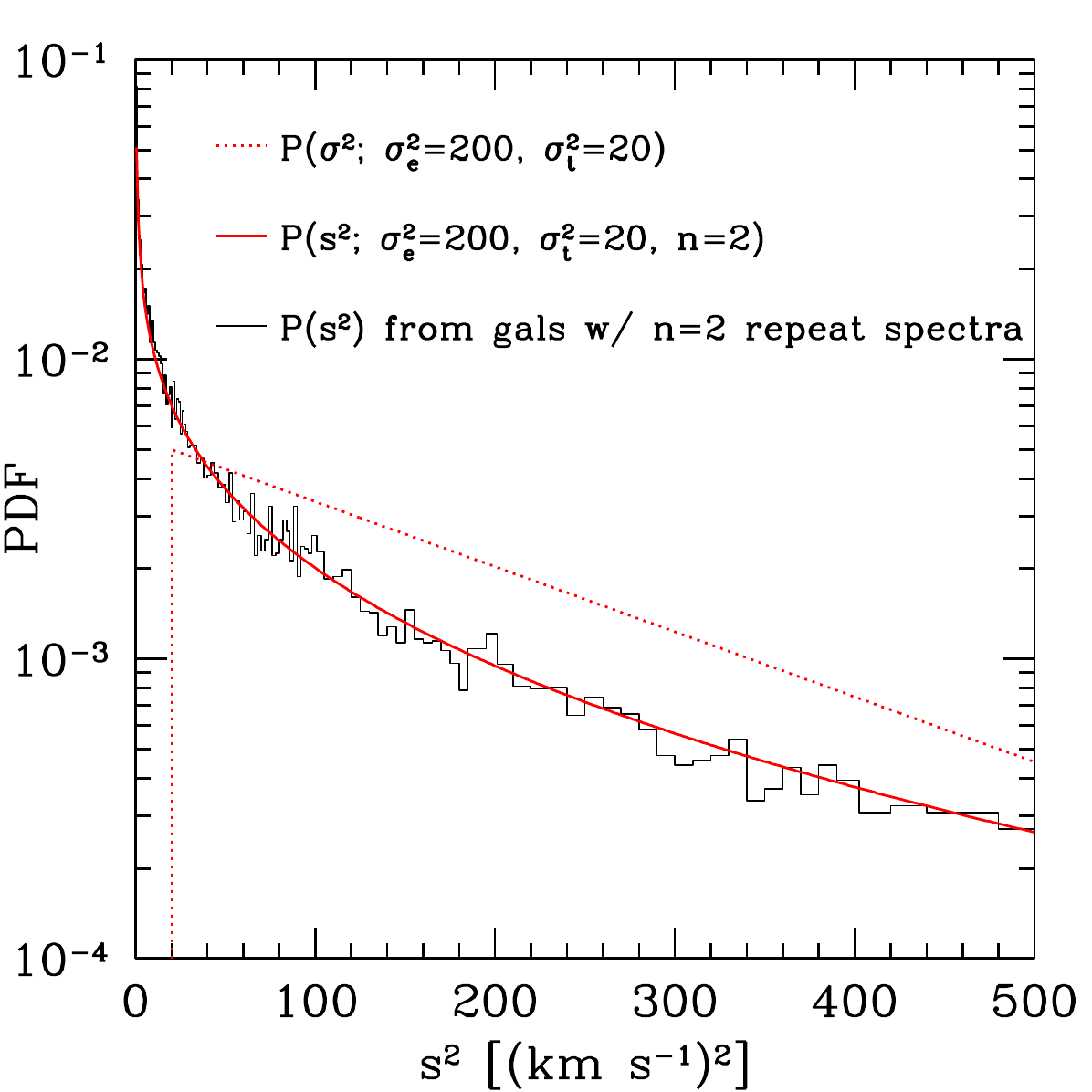}
\caption{
Distribution of sample variance $s^2$ of redshift errors. For each galaxy
with $n=2$ repeat spectra, the sample variance $s^2$ is computed as $s^2=\sum_{i=1}^n
[c(z_i-\bar{z})/(1+\bar{z})]^2/(n-1)$. The distribution is consistent
with the expected $s^2$ distribution (solid curve) for the following case:
the redshift error of each galaxies follows a Gaussian distribution, while the
variance $\sigma^2$ of the Gaussian distribution varies from galaxy to galaxy
and this stochastic variance $\sigma^2$ follows an exponential distribution
with a cutoff (illustrated by the dotted curve, characterized by a scale
parameter $\sigma_e^2$ and a cutoff threshold $\sigma_t^2$).
The description provides an explanation for the redshift error distribution of
SDSS galaxies (Fig.~\ref{fig:zerr}). See text for more details.
} \label{fig:s2stat}
\end{figure}

Why does the redshift error follow more closely to a Laplace distribution
than a Gaussian distribution? Laplace distribution can be thought as the
distribution of Gaussian random variables with mean zero and stochastic
variance that has an exponential distribution. We speculate
that for a given galaxy, the redshift error follows a Gaussian distribution,
but the variance of the distribution varies from galaxy to galaxy. We expect
that for higher variance in redshift errors, there are fewer galaxies, and
that there is a minimum variance (e.g. from the spectral resolution). With
such an expectation, the probability distribution of the variance can be
approximated as an exponential distribution with a cutoff towards lower
value. The overall distribution of the redshift errors then follows a Laplace
distribution with the central part modified by the lower cutoff on the
variance. In such a scenario, the random variable $\Delta v$ can be obtained
through the product of two independent random numbers, $\Delta v=\sigma u$,
where $u$ follows the unit normal distribution and $\sigma^2$ follows a
modified exponential distribution of scale parameter $\sigma_e^2$ with
non-zero values above a threshold $\sigma_t^2$. The two parameters $\sigma_e$
and $\sigma_t$ characterize the tail and central part of the distribution.
The above speculation is supported by the data. In Fig.~\ref{fig:s2stat}, we
show the histogram of the sample variance $s^2$ of the redshift error
distribution estimated from galaxies with $n=2$ repeat spectra, with
$s^2=\sum_{i=1}^n [c(z_i-\bar{z})/(1+\bar{z})]^2/(n-1)$ for each galaxy. The
dotted curve shows an exponential distribution of the variance $\sigma^2$ of
the Gaussian error distribution with $\sigma_e^2=200 {\rm (km\, s^{-1})^2}$
and $\sigma_t^2=20 {\rm (km\, s^{-1})^2}$. The expected distribution of
sample variance $s^2$ is shown as the solid curve. It is derived from
convolving the dotted curve with the sample variance distribution at given
$\sigma^2$ [noting that $(n-1)s^2/\sigma^2$ follows a $\chi^2$ distribution
with $n-1$ degrees of freedom]. Clearly, the $s^2$ distribution from galaxies
with repeat spectra is consistent with being from a sample of Gaussian errors
with stochastic variance that follows a truncated exponential distribution.

In this paper, we adopt the Gaussian-convolved Laplace distribution to model
the redshift error distribution. From fitting the histograms of redshift
errors with such a distribution for different luminosity-threshold galaxy
samples, we derive the parameters $\sigma_{\rm gau}$ and $\sigma_{\rm exp}$.
We find that it is sufficient to adopt two groups of parameters, for bright
and faint galaxies, respectively. We use $\sigma_{\rm gau}=2\kms$ and
$\sigma_{\rm exp}=12.5\kms$ for the luminosity threshold samples of
$M_r<-18$, $-18.5$ and $-19$, while the more luminous samples have
$\sigma_{\rm gau}=5\kms$ and $\sigma_{\rm exp}=14.5\kms$. We incorporate the
redshift errors into our model by adding shifts in the redshift-space
positions of galaxies, following the corresponding Gaussian-convolved Laplace
distribution.

\end{appendix}

\bsp	
\label{lastpage}
\end{document}